\documentclass[a4paper,12pt]{amsart}
\usepackage{amssymb,amsmath,mathrsfs}
\usepackage[usenames,dvipsnames]{xcolor}
\usepackage{hyperref}
\usepackage{tensor}
\usepackage{graphicx}
\usepackage[left=1in,top=1in,right=1in,bottom=1in,headheight=0.5in,foot=0.5in]{geometry}
\usepackage[nodisplayskipstretch]{setspace} 
\usepackage{mdframed}
\usepackage{enumitem}
\usepackage[greek.polutoniko,english]{babel}
\usepackage{appendix}
\usepackage{tikz}
\usepackage{placeins}

\usepackage{float} % For precise float control

\hypersetup{
	colorlinks=true,         
	linkcolor=MidnightBlue,          
	citecolor=MidnightBlue,
	urlcolor=MidnightBlue            
 }
 \let\oldmarginpar\marginpar
\renewcommand\marginpar[1]{\oldmarginpar{\color{red}\raggedright\scriptsize #1}}
\let\oldmarginpar\marginpar
\renewcommand\marginpar[1]{\oldmarginpar{\color{red}\raggedright\scriptsize #1}}

\newcommand{\bra}[1]{\ensuremath{\lf< #1 \rt| }}
\newcommand{\ket}[1]{\ensuremath{\lf| #1 \rt> }}

\newcommand{\comment}[1]{}

\setlist[enumerate,1]{label=\Roman*}
\setlist[enumerate,2]{label=\roman*}

\theoremstyle{definition}

\def\lf {\ensuremath{\left}}
\def\rt {\ensuremath{\right}}

\usepackage{natbib}
\setcitestyle{aysep={}} % author date

\title{\sc Decoherence and Probability}

\usepackage{amsaddr}

\author{}

\author{Richard Dawid}
\address{\vspace{-0.8pc}Stockholm University}
\email{\href{richard.dawid@philosophy.su.se}{richard.dawid@philosophy.su.se}}

\author{Karim P. Y. Th\'ebault}
\address{\vspace{-0.8pc}University of Bristol}
\email{\href{mailto:karim.thebault@bristol.ac.uk}{karim.thebault@bristol.ac.uk}}
\date{\textit{ \today}}
\makeatletter
\let\uppercasenonmath\@gobble

\begin{document}
\setstretch{1.1}
\maketitle

\begin{abstract}
One cannot justifiably presuppose the physical salience of structures derived via decoherence theory based upon an entirely uninterpreted use of the quantum formalism. Non-probabilistic accounts of the emergence of probability via decoherence are unconvincing. An alternative account of the emergence of probability involves the combination of a partially interpreted decoherence model and an averaging of observables with respect to a positive-definite quasi-probability function and neglect of terms $O(\hbar)$. Our analysis delimits the context in which the combination of decoherence and a semi-classical averaging allows us to recover a classical probability model within an emergent coarse-grained description. 
\end{abstract}
\maketitle
\section{Introduction}

The interpretation of probability is a variously contested subject in both philosophy and the foundations of physics. There are, perhaps, two points free from controversy, however. First, \textit{formally} a classical probability can be defined as a \textit{mathematical structure} given by a normalized, positive, and $\sigma$-additive measure over a suitable algebra of events. Second, in using such a structure to \textit{represent} physical states of affairs, we are committing to at least a \textit{partial interpretation}, in the sense of \cite{carnap:1958}, of the measure as (in some sense) a \textit{weighting of possibilities}. Whether such weightings, and such possibilities, should be understood as epistemic or ontic; or, for that matter, subjective or objective, is still left open by such a partial interpretation. Nevertheless, a probability is not purely a mathematical object when we are in the business of physical representation, even absent a full interpretation.  

These remarks prove enlightening when considered in the context of  discussions of probability and emergence in the Many Worlds or Everett approach to the interpretation of quantum theory. In particular, consider the \textit{non-probabilistic} emergentist account of \cite{Wallace:2012}, \cite{Saunders:2021} and \cite{franklin:2023} in which features of the \textit{uninterpreted}  formalism of quantum theory are claimed to be sufficient, in some contexts to some degree, to justify a link between the \textit{Born weightings} and \textit{physical salience}. Decoherence, on this view, is designated a dynamical process under which stable, quasi-classical, `branching structure' emerges. This emergence is intended to be entirely independent of probabilistic interpretive assumptions. The link between the Born weightings and physical salience is established either directly within the quantum formalism via `structural stability' arguments or indirectly based upon similarity between the quantum formalism and interpreted classical physics.  

In Section \ref{probinMW} we argued towards the failure of this approach. First, one cannot justifiably presuppose the physical salience of structures derived via decoherence theory based upon an entirely uninterpreted use of the quantum formalism. Second, one cannot coherently utilise appeal to the limiting classical theory as a requirement for interpretational content, and assert that one is presenting an interpretation of the theory at a quantum level, as opposed to a pragmatic prescription for its application in decohered contexts. Rather, in the context of any putative interpretation of quantum theory, a generalised concept of quantum measure or quantum probability as a weighting of possibilities must be assumed in the application of decoherence theory.  The non-probabilistic emergentist approach to quantum decoherence fails since it denies itself precisely the conceptual resources needed to link Born weightings with physical salience. \textit{Nothing comes from nothing}.

%In this vein, it is worth noting that the probabilities which come into quantum theory via the experimentally confirmed expectation values \textit{cannot} be represented as measures induced by the integration of a genuine probability density function over phase space. Rather, they can, at best, be represented in terms of the marginal probability distributions for position and momentum considered separately, with the density function taking the form of Wigner function, which is formally a \textit{quasi-probability} density function in a \textit{quantum phase space} representation \citep{Wigner:1932,Wigner:1971}. 

In what follows we will construct a novel account of the emergence of probability within the quantum phase space formalism. Our approach takes as its starting point a partially interpreted quantum quasi-probability structure within the quantum phase space approach \citep{Wigner:1932,Wigner:1971}. We draw crucially upon the framing of emergence due to  \cite{butterfield:2011} and \cite{palacios:2022} and of the semi-classical limit due to \cite{feintzeig:2020}.  We show that the combined application of  decoherence and semi-classical averaging, which neglects terms $O(\hbar)$, leads to a classical probability model as an emergent coarse-grained description. This is not, of course, to offer a solution to the measurement problem in terms of a \textit{full interpretation} of the relevant possibility spaces. Rather, what we offer is a conceptual framework for the analysis of classical and quantum probability within which any coherent interpretation of quantum mechanics must be expected to operate. Our aim is to clarify in which sense the principles of quantum mechanics require probabilistic concepts that could be connected to probabilistic measurement outcomes in a cogent way once a consistent interpretation of quantum mechanics has been established. 

The structure of the paper is as follows. Section \ref{probinMW} provides detailed exegesis and rebuttal of the non-probabilistic emergentist account of probability via decoherence focusing on the role of similarity arguments. Section \ref{pandp} then provides a formal reconstruction of classical stochastic and quantum phase space mechanics to allow direct comparison as \textit{classical and quantum possibility space models}. Section \ref{Wemergence} demonstrates the sense in which classical possibility space models can be understood to emerge from quantum possibility space models based upon the \textit{combination} of quasi-probabilistic emergence via decoherence leading to a positive-definite quasi-probability function and coarse-grained emergence via semi-classical averaging of observables which neglect terms $O(\hbar)$. Section \ref{Outlook} recapitulates our results in the context of the motivating dialectic and offers some thoughts regarding outstanding issues of interest. 

%On our analysis, an account of the role of probability in quantum mechanics can most plausibly play out in only one of two ways. First, probability can be introduced as a fully formed classical probability in connection with an extra posit such as collapse, hidden variables, measurement, or observers. Second, one can abstain from extra posits, and establish the probabilistic nature of quantum mechanics as an approximate, emergent concept. In the latter case, there is no plausible way to avoid adding to pure wave mechanics a partial interpretation in terms of possibility weightings. In particular,  there is no way to understand decoherence in general, or the suppression of small amplitudes in particular, absent a partially interpreted structure that weights possibilities. Formally, such weightings can be expected to have the structure of quasi-probabilities (or quasi-measures). On this approach, there are no true classical probabilities at a fine-grained level of description, only quantum probabilities that in some circumstances and to some extent resemble their classical counterparts. 

\section{Emergence and Everett}
\label{probinMW}
\subsection{Decoherence and Emergence}
\label{similarityandemergence}

The role of probability in the interpretation of quantum mechanics takes centre stage in the context of the relationship between the Everett interpretation and decoherence. In particular, according to what might be called the \textit{Oxford  approach} the Born rule can be extracted from the Many Worlds branching structure based on principles of reasoning that leave the application of the Born rule as the only rational way of betting on quantum outcomes open to an agent on an Everettian branch of the wave function who endorses the Everett interpretation.\footnote{See  \cite{Deutsch:1999,Wallace:2002,Wallace:2012,Saunders:2004,Saunders:2005,Wallace:2009,Greaves:2010,Price:2008,Rae:2009,dizadji:2013,Adlam:2014,Dawid:2014,read:2018,brown:2020,steeger:2022,march:2023,saunders:2021b,Short:2023,saunders:2024}.}

\cite{Zurek:2005} and \cite{Baker:2007} point to a circularity in Oxford line of reasoning: decoherence must already be assumed to establish the branching of the wave function that provides the basis for identifying an agent who can consider betting along the lines of the decision theoretic argument. But decoherence already relies on a probabilistic interpretation of the process. In this context, \cite{dawid:2015} diagnose a deeper problem: the notion of probability argued to be required for understanding decoherence in the sense of a probabilistic suppression of off-diagonal elements of the density matrix is taken to be stronger than the decision theoretic notion of probability offered. This approach to extracting the Born rule therefore is not just circular but incoherent.

The original non-probability presentation of decoherence in the Oxford view is due to \cite{Wallace:2010,Wallace:2012}, who suggests that ``[w]e can think of the significance of the Hilbert space metric as telling us when some emergent structure really is robustly present, and when it’s just a `trick of the light' that goes away when we slightly perturb the microphysics...What makes perturbations that are small in Hilbert-space norm ‘slight’, [is] not the probability interpretation of them. Ultimately, the Hilbert- space norm is just a natural measure of state perturbations in Hilbert space, and that naturalness follows from considerations of the microphysical dynamics, independent of higher-level issues of probability'' (pp. 253–254). However, as noted by \cite{dawid:2015}, this purely structural emergentist account provides no \textit{justification} for the connection between Born weightings and physical salience. 

In this context, \cite{Saunders:2021} and \cite{franklin:2023} have sought to further develop Wallace's approach. Franklin argues that ``the neglect of terms with relatively small amplitudes can be justified non-probabilistically [...] in contexts where interference is rife, the probabilistic interpretation of the (mod-squared) amplitudes is ruled out [...] the Born rule, in such contexts, takes the form of an averaging measure rather than a probability measure. [...] we should think of the relation between small amplitudes and irrelevance as a dynamical phenomenon. The relative magnitude of the amplitudes encodes the dynamical contribution of each term.''

Saunders argues using slightly different language towards the same central point. In particular, he claims that ``Strongly peaked amplitude'' does not, prior to defining the branching structure of the state, have to be interpreted as ``highly probable.'' [...] the “average values of local densities” are defined not by averaging the densities, but as the values of the local densities on those trajectories on which the amplitudes are (very sharply) peaked. In the case of Ehrenfest's theorem, whilst it is possible to interpret $\langle x \rangle_\psi$ operationally, in terms of multiple measurements [...] it is also possible to interpret it realistically, as the location of the peak of the wave-function as it evolves over time, in accordance with classical equations...''. 

What Wallace, Franklin and Saunders all seem to have in mind is that features of the evolution equations of quantum theory are sufficient, in some contexts to some degree, to justify a link between the Born weightings and \textit{physical salience}, without these weightings being understood probabilistically in any sense. The justification for this connection is then provided in terms of structural similarity to the classical theory. The problems with such a emergentist view will be considered  in the following section in the context of the reliance on \textit{uninterpreted similarity arguments}. 

\subsection{Similarity and Interpretation}
\label{mwintrouble}

The failure of non-probabilistic emergence based on uninterpreted similarity arguments can be understood to arise from a basic conflict with the principle that a scientific theory should allow for empirical testing on its own terms. The key problem is the assumption that the set of rules that specify an important part of the theory's empirical import, namely the decoherence of branches of the wave function, can be extracted from observing structural similarities to a theory that serves as a limiting case of that theory -- the model where coherence terms are set to zero. In other words, a `limiting theory' serves as the basis for extracting empirical implications of the fundamental theory.\footnote{Here and below we assume the `factual' understanding limits as per Section \S\ref{Emergence}.}  

The problem with this line of reasoning is that it does not explain what measuring a certain value of an observable \textit{means} at the level of the full theory. As long as no such understanding is forthcoming at the level of the full theory, however, \textit{we have no basis to decide whether or not we are justified to call any other theory a limiting theory of our full theory}. Mere similarity arguments are insufficient for making that decision  for one principal reason: as long as the implications of measurements cannot be spelled out at the level of the full theory, we remain insensitive to the distinction between empirically relevant stable dynamics on the one hand and spurious dynamics of parameterization prescriptions on the other. In the limiting theory that sets coherence effects to zero, the set of allowed states are confined to states that show no coherence effects. Any discovery of coherence effects would therefore contradict the limiting theory. The question as to whether coherence should be considered probable or improbable thus does not arise. Coherence is ruled out. In the full theory, coherence is consistent with the theory. Coherence effects are represented in the theory's set of allowed states. To understand whether they are suppressed or not, it is not sufficient to point at  a small dimensionless number that characterizes cohered states because small dimensionless numbers might, in principle, also be extracted from specific parameterizations of the theory that bear no physical significance. To rule out this possibility, one needs to find the basis for those states at the level of the full theory. Such an analysis \textit{requires} appeal to a generalised concept of quantum measure or quantum probability as a weighting of possibilities.

The similarity approach has a second, related problem. While a limiting theory can be deduced from a fundamental theory, the opposite is not true. A probabilistic interpretation of the limiting theory with zero coherence (to the extent it can be given) does not formally imply the probabilistic characteristics of the fully quantum regime in terms of the full Born rule. In other words, we end up deploying two entirely different lines of reasoning to establish what formally looks like one coherent concept of quantum probability.  All this is a far cry from the initial claim that Many Worlds quantum mechanics has the attractive feature to require no posits beyond the wave function equations. Indeed, an appeal to decoherence as a precondition of any interpretational content would render the Many Worlds approach of a piece with precisely the pragmatic, neo-Bohrian outlook that is inconsistent for any claim of \textit{realism about the quantum state}, whether presented in functionalist/structuralist terms or otherwise.\footnote{See \cite[\S1.8]{Wallace:2012}. What we say here is perfectly compatible with the (very reasonable) view that the `worlds' in Many Worlds require decoherence to be attributed interpretational content.}  For example, such an approach would involve implementing the precisely the prescription on the use of the Born rule as a probabilistic rule advocated by \cite{healey:2017}. Pragmatic approaches to quantum theory are without doubt interesting and valuable in their own rights. However, we do not take a marriage with the Many Worlds view of quantum mechanics to be a union that would be to the profit of either party.

Viewing the similarity argument from a slightly different angle may contribute to understanding both the reason for its intuitive appeal and the point where it goes wrong. It is, of course, striking that the decohered limit of wave mechanics looks so similar to a model with a classical probability function. If quantum mechanics were new, no probabilistic interpretation of the mod-squared amplitudes were known, and there were no understanding of the theory's empirical implications, it would be plausible to infer from the stated similarity argument alone that quantum theory most probably has an interpretation that allows for neglecting small amplitudes. The similarity just looks too nice to be accidental. Heuristic reasoning of this kind is standard fare in physics and is often successful, which explains its intuitive appeal. Indeed, as a strategy of grasping how a theory works and assessing the chances that a theory is viable, this kind of reasoning is perfectly legitimate.  

It is crucial to understand, however, that the foundational debate does not play out at the described epistemic level. Physicists have accepted quantum theory as the viable theory of microphysics for one hundred years. What is at stake in the foundational debate is not the assessment of a theory's viability in the absence of a full understanding of that theory. What is at stake is the full understanding. And when it comes to fully spelling out the theory,  no structural similarity argument with another theory can take over that conceptual task. 

In summary, returning to our principal argument: As long as no probabilistic interpretation of the wave function is provided at the level of quantum mechanics, it is not clear whether Born weights are a characteristic of physically relevant dynamics or of a spurious parameterization. Therefore, it is not justified to infer the empirical import of Many Worlds quantum mechanics from the fact that the resulting wave function in a given limit looks strikingly similar to the empirical results of a zero coherence theory. One might assert by fiat that the import of Many Worlds quantum mechanics matches the import of the corresponding zero coherence theory in a given limit. If one goes down that road, however, the zero coherence theory turns from a limiting theory of the full quantum theory into an essential element of quantum theory that is needed for providing the link between the theory's formal structure and its empirical import. The result is a confusing compound of mutually dependent theoretical posits. We cannot make valid inferences about the world based upon uninterpreted similarity arguments combined with the formal structure of a decoherence model. 

\section{Probability and Possibility}
\label{pandp}

The previous section demonstrated that a probabilistic understanding of quantum mechanics needs to be established at the level of the full theory. This conclusion stands in conflict, however, with a  second step of reasoning put forward by  \cite{franklin:2023}.  
Franklin writes: ``[A] probabilistic interpretation of the mod-squared amplitudes is inapplicable before decoherence has occurred. In the presence of interference amplitudes may cancel each other out -- thus, interpreting amplitudes in such contexts probabilistically will not do. It is only when interference is sufficiently suppressed that mod-squared amplitudes approximately conform to the probability axioms: any attempt to interpret mod-squared amplitudes as probabilities in the presence of interference will be empirically undermined [...].  Therefore, at least in some of the contexts where the Born rule measure is applied and expectation values are discussed these are not to be given a probabilistic interpretation." \citep[pp. 13-14]{franklin:2023}. 

Franklin thus argues that it is misguided to even look for a probabilistic interpretation of the dynamics at the quantum level because quantum theory provides no basis for a quantum probability measure that satisfies the Kolmogorov axioms.\footnote{See \citep{fine:1982,fine:1982b,pitowsky:1989,suppes:1993,Hartmann:2015,Wallace:2014}.} On his reasoning, \cite{Zurek:2005}, \cite{Baker:2007} and \cite{dawid:2015} are not just wrong in claiming that decoherence \textit{needs} to be based on a probabilistic interpretation of quantum processes. They are already wrong in assuming that a probabilistic interpretation of the quantum regime is a meaningful goal. Franklin asserts that establishing a probabilistic account at the level of the limiting theory is the \textit{only} way to get from quantum mechanics to empirical predictions. In this section and the next we will carry out a detailed analysis of this issue and present a proposal for the precise sense in which probabilistic concepts can be identified at the level of the full theory. 

Before commencing this detailed analysis, it is helpful to spell out the goal of our investigation. A full understanding of the way the probabilistic characteristics of quantum measurements link back to features of quantum mechanics would require a viable interpretation of quantum mechanics. Since the present analysis does not propose an interpretation of quantum mechanics, the goal of our analysis must be more modest. Our aim is to clarify in which sense the principles of quantum mechanics require probabilistic concepts that could be connected to probabilistic measurement outcomes in a cogent way once a consistent interpretation of quantum mechanics has been established. 

Although the connection between the basic probabilistic concepts of quantum mechanics and probabilistic empirical predictions can be expected to play out in different ways -- for example, in spontaneous collapse models as opposed to many worlds interpretations -- we take our analysis of Section 2 to establish, that the identification of the basic probabilistic concepts of quantum mechanics at a fundamental level would need to provide the foundation for \textit{any} non-instrumentalist interpretation of quantum mechanics that may be developed. A  strictly instrumentalist interpretation would allow us to introduce a prescription of probabilistic data analysis ``by hand" without linking it to any feature of the theory. In contrast, in the context of a non-instrumentalist interpretation, some generalised probabilistic concepts must be identifiable at the fundamental level of quantum mechanics for the interpretation to make sense of coherent and decoherent phenomena. Our goal will be to provide a comprehensive analysis of such generalised probabilistic concepts. Our analysis remains neutral as to which interpretation they would be applied but is based on the assumption that some interpretation is required.  
 
The following Section \ref{Wemergence} focuses on the physical details. The present section is more formal and involves successively introducing three types of objects: 1) uninterpreted \textit{probability structures}; 2) partial interpretation of these in terms of \textit{possibility weightings}; 3) explicit representation of such a partial interpretation in terms of \textit{possibility space models}.

\subsection{Probability Structures}

\subsubsection{Quasi-Probability Structures}

A quasi-probability structure is a triple $(\Omega,\mathfrak{E},\tilde{\mu})$ where the three elements are defined as follows:
\begin{enumerate}
\item \textbf{Sample Space}: $\Omega$ is a non-empty set;
\item \textbf{Event Algebra}: $\mathfrak{E}$ is a non-empty collection of subsets of $\Omega$ such that:
\begin{enumerate}
\item $\alpha^c\in \mathfrak{E}$ for all $\alpha\in\mathfrak{E}$ (closed under complementation);
\item $\alpha \cup \beta\in \mathfrak{E}$ for all $\alpha,\beta\in\mathfrak{E} $ (closed under finite union);
\end{enumerate}
\item  \textbf{Quasi-Measure}: $\tilde{\mu}$ is a set function $\tilde{\mu}: \mathfrak{E}\rightarrow \mathbb{R}$ which is such that $\tilde{\mu}(\Omega)=1$ (normalized) and $\tilde{\mu}(\emptyset)=0$ (empty set is measure zero).
\end{enumerate}
By definition we have that $\emptyset\in\mathfrak{E}$, $\Omega\in\mathfrak{E}$, $\mathfrak{E}$ is closed under finite intersection, and $\alpha^c=  \Omega\setminus \alpha $

Two important features that a quasi-probability model \textit{need not} have are $\sigma$-additivity and positivity. The first is since we have not insisted that the event algebra $\mathfrak{E}$ is a $\sigma$-algebra; it need not be closed under countable unions, cf. \citep{Arageorgis:2017}. The second is since we have not insisted that the quasi-measure $\tilde{\mu}$ is a measure; it need not be positive (nor indeed $\sigma$-additive). Strengthening the model to include both these features results in the familiar formal structure of a classical probability model. 

\subsubsection{Classical Probability Structures}

A classical probability structure is a triple $(\Omega,\Sigma,\mu)$ where the three elements are defined as follows:
\begin{enumerate}\setcounter{enumi}{3}
\item \textbf{Sample Space}: $\Omega$ is a non-empty set;
\item \textbf{$\sigma$-Algebra}: $\Sigma$ is a non-empty collection of subsets of $\Omega$ such that:
\begin{enumerate}
\item $\sigma^c\in\Sigma$ for all $\sigma\in\Sigma$ (closed under complementation);
\item $\sigma_1 \cup \sigma_2 \cup \sigma_3 ...  \in \Sigma$ for all $\sigma_1, \sigma_2, \sigma_3 ...  \in \Sigma$ (closed under countable union);
\end{enumerate}
\item  \textbf{Probability Measure}\label{pmeasure}: $\mu$ is a set function $\mu: \Sigma
\rightarrow \mathbb{R}$ such that:
\begin{enumerate}
\item $\mu(\Omega)=1$ (normalized)
\item $\mu(\sigma)\geq 0$ for all $\sigma\in\Sigma$ (positive)
\item\label{sigad} $\mu( \sigma_1 \cup \sigma_2 \cup \sigma_3 ...) = \mu(\sigma_1)+ \mu(\sigma_2)+\mu(\sigma_3) ...$ for a countable collection of mutually disjoint algebra elements $\sigma_1, \sigma_2, \sigma_3 ...  \in \Sigma$ ($\sigma$-additivite).
\end{enumerate}
\end{enumerate}

Additivity means that probability measures are automatically such that $\mu(\emptyset)=0$. Clearly, every classical probability model is a quasi-probability model. By design, the probability measure in a classical probability model will satisfy the Kolmogorov probability axioms.

\subsection{Partial Interpretation via Possibility Weightings}

Here will consider the general sense in which probability structures can be \textit{partially interpreted} to provide \textit{possibility weightings}. Following \cite{carnap:1958}, a partial interpretation is an assignment of meaning to theoretical structure such that there is a range of admissible interpretations in the complete language. A partial interpretation thus allows for the interpretation of theoretical structure to be strengthened by further postulates \citep{suppe:1971,sep-theoretical-terms-science}. 

In the context of a probability structure a partial interpretation in terms of a possibility weighting is an assignment of meaning to the quasi-measure or measure such that there is a range of admissible interpretations in terms of a full interpretation of probability or quasi-probability. In general terms, such meaning is tied to the conceptualisation of the measure or quasi-measure as a \textit{weighting of possibility}. With respect to a quasi-measure $\tilde{\mu}$ (each of these holds mutatis mutandis for measure $\mu$) the four basic properties of such a weighting are as follows: Any event, $\alpha$, for which $|\tilde{\mu}(\alpha)|\neq0$ is \textit{possible}. Any event, $\alpha$, for which $|\tilde{\mu}(\alpha)|=0$ is \textit{impossible}. Any event, $\alpha$, for which $|\tilde{\mu}(\alpha)|=1$ is \textit{certain}.  For any two events  $\alpha$ and $\beta$ the event $\alpha$ has a \textit{higher possibility weight} than the event $\beta$ if and only if $|\tilde{\mu}(\alpha)|>|\tilde{\mu}(\beta)|$ where the \textit{quantitive difference} in the possibility weighting is given by $\delta=|\tilde{\mu}(\alpha)|-|\tilde{\mu}(\beta)|$.

The concepts `possible', `impossible', and `certain' can be given their natural linguistic meaning. The crucial concept of `higher possibility weight', by contrast, requires some further interpretation to be fully understood. We do not intend to provide this here, but rather set-out the framework in which any such interpretation must work. An event being higher weighted than another event indicates that it is \textit{quantitatively closer} to certain events and further from impossible events, with the closeness given by the quasi-measure. However, the concept of higher possibility weighting is a \textit{only a partial interpretation} of the  quasi-measure structure in that it is an assignment of meaning to a structure which allows for a range of further admissible interpretation. In the case of classical probability and measures, such further admissible interpretations can be directly provided in terms of standard notions of credence, frequency, or physical chance. In the case of a quasi-probability the structure of admissible interpretations are required to be more exotic or include further ingredients. We will return to this issue in our closing remarks. 

\subsection{Classical and Quantum Possibility Space Models}

\subsubsection{Classical Possibility Space Models}

A classical possibility space model is a triple $(\Gamma,\mathfrak{O},\rho)$ that takes the following form:
\begin{enumerate}\setcounter{enumi}{6}
\item \textbf{State Space}\label{SS}: $\Gamma=\mathbb{R}^{2N}$ represents the space of possible states of system as a  $2N$-dimensional symplectic manifold equipped with the closed non-degenerate two form $\omega=dq\wedge dp$ and associated volume measure $dq\cdot dp$ in the Darboux chart;
\item \textbf{Observable Algebra}\label{OA}: $\mathfrak{O}$ represents observables as a Poisson algebra given by the space of real-valued smooth functions over $\Gamma$ with the Cartesian product $\cdot$ and Poisson bracket $\{,\}$, the relevant bilinear products. The distinguished function $H\in\mathfrak{O}$ induces a time evolution automorphism via the Poisson bracket: $\frac{d}{dt} A = \{A,H\}$ for all $A\in\mathfrak{O}$.
\item  \textbf{Probability Density Function}\label{PDF}: $\rho$ is a phase space probability density function, $\rho(q,p): \Gamma
\rightarrow \mathbb{R}$, which is Lebesgue integrable with respect to the volume measure, $dq\cdot dp$, and induces a probability measure, $\mu(B)=\int_{B} \rho(q,p) dq\cdot dp$ that satisfies the conditions: 
\begin{enumerate}
\item $ \mu(B)\geq0$ for all $B \in\mathcal{B}$ (positive) 
 \item $\int_{\Gamma} \rho(q,p) dq\cdot dp  =1$ (normalized)
\item \label{sadd} If $B_1,...,B_n,...\in\mathcal{B} \: \text{with}\:B_i\cap B_j =\emptyset \:\text{for}\: i\neq j$
then $\mu(\cup_{n=1}^\infty B_n)=\sum_{n=1}^\infty \int_{B_n} \rho(q,p) dq\cdot dp$ ($\sigma$-additive)
\end{enumerate}
where $B\in\mathcal{B}$ are the Borel sets $\mathcal{B}(\mathbb{R}^{2N})$. 
\item \textbf{Expectation Values}\label{EV}: $\langle A \rangle$ is the expectation value or mean of an observable defined as: $\langle A \rangle 
\equiv \int_\Gamma A(q,p) \cdot \rho(q,p) dq \cdot dp$ for all $A\in\mathfrak{O}$
\end{enumerate}

The conditions on the representation \ref{SS}--\ref{EV} encode two features which will be important for the comparison with phase space representations of quantum possibly spaces. These are the \textit{local conservation} and \textit{localisability} of probability density.  

The local conservation of probability density is a well known feature of a phase space representations of a classical possibly model. It is typically expressed via the Liouville equation: 
\begin{equation}
\label{Leq}
\frac{d \rho}{ d t} = \frac{\partial \rho}{\partial t} + \{\rho,H\} = 0
\end{equation} 
This equation guarantees that the additivity property of regions of phase space is locally preserved \textit{over time}. If we think of the probability like a fluid, we can understand there to be a phase space 3-current given by the tuple $(\rho,\rho\dot{q},\rho\dot{p})$. The Liouville equation then expresses the fact that the current represents an \textit{incompressible flow} and is a result of the absence of compression or rarefaction points in the probability `fluid'  \cite[p. 28]{pathria:2011}, cf. \cite[p.11]{gibbs:1902} and the density about the representative point in the flow is conserved.

The localisability of probability density is much less discussed but will be equally important for our discussions. The phase space representation given by conditions \ref{SS}--\ref{EV} is such that the \textit{essential support} of the probability density function $\rho(q,p)$ is given by phase space points $\{q,p\}$. The essential support of a function, $\mathtt{ess\,sup}(f)$, is the smallest closed subset in the domain of a measurable function such that the function can be zero `almost' everywhere outside that subset. The `almost' in this context is cashed out via the measure such that the points which are outside the essential support and where the function is non-zero are of measure zero. For any Lebesgue measurable function $f$ we have that $\mathtt{ess\,sup}(f)=\mathtt{sup}(f)$ \citep[p.13]{lieb:2001}. The important feature to hold in mind for our discussion is that essential support (and support) of $\rho(q,p)$ is given by the smallest possible phase space regions such that the function can be zero (almost) everywhere else. These are phase space points (the singleton elements of the Borel sets). This means that it is possible to consider probability density functions that are (almost) entirely concentrated at a single point which amounts to allowing the possibility that the probability density function approximates a $\delta$-function.  Correspondingly, since its integral over phase space is normalised, by concentrating a probability density function almost entirely at one point we must allow that the function is unbounded from above. 

A stochastic phase space model provides a partial interpretation of a classical probabilistic structure. The state space $\Gamma$ is the sample space $\Omega$. The Borel sets given by sub-regions of phase space $\mathcal{B}(\mathbb{R}^{2N})$ are the $\sigma$-algebra \citep{feller:1991}. The probability measure $\mu(B)$ is given by the integration of the probability density function $\rho(q,p)$ with respect to the volume measure $dq\cdot dp$ over a sub-region $B\subseteq \mathbb{R}^{2N}$. 
The connection between the  conditions \ref{PDF}\ref{sadd} and 
\ref{pmeasure}\ref{sigad} is guaranteed by the definition of $\sigma$-algebra. The event corresponding to the entire space is certain. The event corresponding to the empty set is impossible. The probability measure provides us a representations of events being possible and being higher weighted. The model includes a deterministic subset since a function that approximates a $\delta$-function is an admissible PDF and thus the case in which the singleton of the Borel sets is measure one and all other points are measure zero is an admissible stochastic phase space model. It is worth emphasising that a stochastic phase space model is not a full interpretation of a classical probabilistic structure since the weighting of possibility provided in terms of the probability measure admits further admissible interpretation in terms of standard notions of credence, frequency, or physical chance.

\subsubsection{Quantum Possibility Space Models}
\label{QOSM}

A quantum possibility space model is a triple $(\Gamma,\mathcal{O},F)$ that takes the following form:
\begin{enumerate}\setcounter{enumi}{10}
\item \textbf{State Space}\label{QSS}: $\Gamma=\mathbb{R}^{2N}$ represents the space of possible states of system as a $2N$-dimensional symplectic manifold equipped with the closed non-degenerate two form $\omega=dq\wedge dp$ and associated volume measure $dq\cdot dp$ in the Darboux chart;
\item \textbf{Observable Algebra}\label{QOA}: $\mathfrak{A}$ represents observables as a (non-commutative) Moyal algebra of Weyl symbols which are the Wigner transforms of the algebra of (Weyl ordered) bounded linear operators $\mathcal{B}(\mathcal{H})$ on a Hilbert space of square integral functions $\mathcal{H}=L^2(\mathbb{R}^{2N})$. The binary operation is given by a $\star$-product operation which can be expressed as a pseudo-differential operator in powers of $\hbar$ and the non-commutativity of the algebra is expressed via the fundamental relation that  $[\hat{A},\hat{B}]=\{\{A,B\}\}\equiv\frac{1}{i\hbar}(A\star B -A \star B)$ for all $A,B\in\mathfrak{A}$ and all $\hat{A},\hat{B}\in\mathcal{B}(\mathcal{H})$. The distinguished function $H\in\mathfrak{A}$ induces a time evolution automorphism via the Moyal bracket such that $\frac{d}{dt} A = \{\{A,H\}\}$ for all $A\in\mathfrak{A}$;
\item  \textbf{Quasi-Probability Density Function}\label{QPDF}:  is a possibility space weighting function $F(q,p): \Gamma
\rightarrow \mathbb{R}$ that induces a quasi-measure $\tilde{\mu}(B) =\int_B F(q,p)dq\cdot dp$ that satisfies the conditions: 
\begin{enumerate}
\item $\tilde{\mu}(\Gamma)=\lim_{n\rightarrow\infty} \int_{\mathbf{B}_n} F(q,p) dq\cdot dp  =1$ where $\mathbf{B}_n = \{ (q,p) \mid |q|^2+|p|^2 \leq r_n\}$ (normalized)
\item $\mid F(q,p) \mid \leq \frac{1}{\epsilon}$ (bounded)
\item \label{sadd} If $B_1,...,B_n,...\in\mathcal{B} \: \text{with}\:B_i\cap B_j =\emptyset \:\text{for}\: i\neq j$
then $\tilde{\mu}(\cup_{n=1}^\infty B_n)=\sum_{n=1}^\infty \int_{B_n} F(q,p) dq\cdot dp$  ($\sigma$-additive)
\end{enumerate}
where $B\in\mathcal{B}$ are the Borel sets $\mathcal{B}(\mathbb{R}^{2N})$\item \textbf{Expectation Values}\label{QEV}: $\langle A \rangle$ is the expectation value or mean of an observable defined as: $\langle A \rangle 
\equiv \int_\Gamma A(q,p) \star F(q,p) dq \cdot dp$ for all $A\in\mathfrak{A}$.
\end{enumerate}

The contrast between classical and quantum possibly space models is greatly clarified by examining the \textit{failure} of \textit{local conservation} and \textit{localisability}  of quasi-probability density implied by the conditions \ref{QSS}--\ref{QEV}. Let us demonstrate this failure explicitly for the choice of Wigner function, $W$ as the quasi-probability density function \citep{curtright:2013}.

Failure of local conservation of quasi-probability is a direct consequence of the non-commutativity of the Moyal algebra of quantum phase space observables in comparison to the Poisson algebra of classical phase space observables as encoded in the relation $\{\{A,B\}\}=\{A,B\}+O(\hbar)$.  We can show this explicitly by considering the quasi-probability flux for some arbitrary region of phase space $\mathcal{S} $. This is given by the expression \cite[p. 57]{curtright:2013}:
\begin{eqnarray}
\label{qefflux}
\frac{d}{dt}\int_\mathcal{S} dq dp  W  &=& \int_\mathcal{S} dq dp \Big{(} \frac{\partial W}{\partial t} +\dot{q}\frac{\partial W}{\partial q}+\dot{p}\frac{\partial W}{\partial p}\Big{)}  \\
&=&  \int_\mathcal{S} dq dp \Big{(}  \{\{H, W\}\} -  \{H,W\}  \Big{)} \nonumber \\
&=&O(\hbar) \nonumber
\end{eqnarray} 
where we have used the Wigner transform of the Heisenberg equations of motion $\dot{q}= \frac{\partial H}{\partial p}$ and $\dot{p}= -\frac{\partial H}{\partial q}$ and the Moyal equation $\frac{d}{dt} W = \{\{H,W\}\}$. The quasi-probability density associated with regions of phase space thus manifests a violation of \textit{local additivity over time} in terms of failure of conservation of the probability density in the precise sense of the density about the representative point in the flow not being conserved, cf. \cite[p. 58]{curtright:2013}. This is despite retaining satisfying additivity  \textit{at a time} in terms of $\sigma$-additivity. The failure of local conservation in this sense is in marked contrast to the classical probability density function in phase space, cf. \cite[p.23]{Wallace:2021}.

The failure of localisability can be understood as follows. Although a quasi-probability function need not in general be Lebesgue integrable over the entire phase space \citep{aniello:2016} one can show that, for example, the Wigner function is an element of $L^2(T^\star\mathbb{R}) \cap C_0(T^\star\mathbb{R})$ and is thus a square integrable and continuous function on the phase space \cite[p. 142]{landsman:2012}. Moreover, the induced quasi-measure is $\sigma$-additive and the Wigner function induces a \textit{finite signed measure} \citep{dias:2019}. It might seem therefore that besides the conservation failure and the negativity the Wigner function is a very much like a classical probability function. The bound, however, leads to an important difference in the types of possibility space representations that a quasi-probability function can provide. By the Cauchy–Schwarz inequality the function is bounded such that $-\frac{2}{\hbar}\leq W(q,p) \leq \frac{2}{\hbar}$ and we thus have that $\epsilon = \frac{\hbar}{2}$. This, in turn, leads to a restriction of $\mathtt{ess\,sup}(W)$ to volumes of phase space greater than equal to one in units of $\hbar$ \citep[p.19]{delllectures}. Thus, in contrast to the classical case, it is \textit{not} possible to concentrate quasi-probability density almost entirely at a single point. This amounts to precluding the possibility that the quasi-probability density function approximates a $\delta$-function in phase space \citep[p.71]{leonhardt:2010}. Phase space points are not in $\mathtt{ess\,sup}(W)$ and we cannot have a situation in which the Wigner function is non-zero at a point but zero (almost) everywhere else. Physically speaking, non-localisablity can be understood as a consequence of the Heisenberg uncertainty principle which, in turn, is a direct consequence of the non-commutative structure induced by the $\star$-product. See \cite[\S5]{curtright:2013} and \cite[\S5.1]{huggett:2021}. The structure of a quantum possibility space model encodes the fact that quantum possibilities are not `distinct' in the sense that the failure of localisability implies we can only `peak' the Wigner function on regions of finite size to a limited extent. This is in addition to the failure of `distinctness' of possibilities via  
the existence of entanglement which is encoded in the negativity of the Wigner function \citep{dahl:2006}. 

A quantum possibility space model provides a partial interpretation of a quasi-probability structure. The state space $\Gamma$ is the sample space $\Omega$ and the event algebra $\mathfrak{E}$ is given by Borel defined by sub-regions of phase space $\mathcal{B}(\mathbb{R}^{2N})$. The quasi-measure $\tilde{\mu}$ is given by the integral of the quasi-probability density function with respect to the volume measure. The induced quasi-measure is a finite signed measure over a $\sigma$-algebra, and thus the probability structure is stronger than that required by a quasi-measure structure as we have defined it. 

\section{Probability and Classicality}
\label{Wemergence}

In this section we demonstrate that the combination of `quasi-probabilistic emergence' and `semi-classical emergence' allow us to derive a classical possibility space model from a quantum possibility space model. This demonstration depends upon two important results. First, that explicit models of decoherence in quantum phase space display the generic feature that the Wigner quasi-probability distribution is positive-definite after finite times of the order of the decoherence time. This is the quasi-probabilistic emergence with Wigner positivity the relevant novel and robust behaviour. Second, that the generalised Ehrenfest relations imply that the classical and quantum moment evolution equations are syntactically isomorphic with the Wigner function with the latter playing the same role as the probability density function in the former. A positive-definite Wigner function then displays localisation and local conservation behaviour identical to that of a probability density function to the extent to which we can neglect terms $O(\hbar)$. This is the semi-classical emergence with localisation and conservation of the relevant novel and robust behaviour. 

\subsection{Wigner Negativity and Decoherence}
\label{WNandDe}

The Wigner function is the central object in the phase space approach to quantum mechanics.\footnote{Physics references are \cite{OConnell:1981,hillery:1984,case:2008,Gosson:2017,leonhardt:2010,curtright:2013}. See \cite{suppes:1961,cohen:1966,sneed:1970,friederich:2021,Wallace:2021} for philosophical analysis.} 
Representing the quantum state of a system via a density matrix, $\hat{\rho}$, the Wigner function, $W(q,p)$, takes the form:
\begin{equation}
W(q,p)=\frac{1}{2\pi \hbar} \int dq'  \bra{ q-q'} \hat{\rho} \ket{ q+q'} e^{-iq'p/\hbar}
\end{equation}
The transformation between the density matrix $\hat{\rho}$ and the Wigner function $W$ can be generalised to an arbitrary operator $\hat{A}$ as:
\begin{equation}
A(q,p)=\frac{1}{2\pi\hbar} \int dq'  \langle q-q'\mid \hat{A} \mid q+q' \rangle e^{-iq'p/\hbar}
\end{equation}
where the \textit{Weyl symbol} $A(q,p)\in\mathfrak{A}$ is the \textit{Wigner transform} of the bounded Hilbert space operator $\hat{A}\in \mathcal{B}(\mathcal{H})$.

An important property of the Wigner transform is that the trace of the product of two operators $\hat{A}$ and $\hat{B}$ is expressed in phase space in terms of the integral of the product of the relevant Wigner transforms:
\begin{equation}
\text{Tr}[\hat{A}\hat{B}]=\frac{1}{\hbar} \int \int A(q,p)B(q,p) dq dp 
\end{equation}
This immediately implies that we can express the expectation value of an operator as:
\begin{equation}
\langle A \rangle = \text{Tr}[\hat{\rho}\hat{A}]= \frac{1}{\hbar}\int \int W(q,p) A(q,p) dq dp
\end{equation}
The Wigner function behaves like a density in that we obtain the average value of a quantity by integrating over that quantity multiplied by the Wigner function.

The Wigner function reproduces the marginal probability densities for position and momentum given by the mod-squared amplitude since we have that: 
\begin{eqnarray}
\mu(q) = \int W(q,p) dp = \bra{q} \hat{\rho} \ket{q} \\
\mu(p) = \int W(q,p) dq = \bra{p} \hat{\rho} \ket{p} 
\end{eqnarray}
It can be proved that any quasi-probability distribution function of the form $F(q,p)=\bra{\psi}  \hat{A}(q,p)\ket{\psi}$ which reproduces the marginal probability densities corresponding to the Born rule cannot also be positive semi-definite \citep{Wigner:1971}. The Wigner function thus cannot be positive semi-definite. Wigner negativity has been variously recognised as the distinctive non-classical feature of the Wigner function and has been shown to have direct implications for both contextually and entanglement \citep{dahl:2006,delfosse:2017,booth:2022}. The size of the regions of negativity in phase space are of order $\hbar$. The subset of Wigner functions that correspond to minimum uncertainty coherent states can be shown to be everywhere positive (and vice versa) \citep{hudson:1974,marino:2021}. 

Despite its negativity, the Wigner function has a number of attractive features that mark it out as privileged among the quasi-probability distribution functions. In particular, the density and marginal features noted above crucially depend upon the $\star$-product associated to the Wigner function being the Moyal $\star$-product. This is what allows one $\star$-product to be dropped inside an integral via integration by parts, leading to formal behaviour that matches that of a genuine probability density function for the marginals and expectation values. This feature is in contrast to other $\star$-product association rules and quasi-probability distributions (such as the Husimi Q-function) which do not reproduce the full set of marginal distributions corresponding to the Born rule \cite[\S13]{curtright:2013}.\footnote{For further discussion see 
\cite{leonhardt:2010,schroeck:2013,friederich:2021,stoica:2021,umekawa:2024}.}

Let us now consider the behaviour of the Wigner function within a simple model of decoherence with a focus on the role of Wigner negativity.  The general framework for the study of decoherence is quantum master equations for the \textit{reduced} density matrix of a quantum system. For our purposes it will suffice to consider the most basic master equation, that due to \cite{joos:1985}. The Joos-Zeh equation can be derived based on an idealised decoherence model with recoilless scattering that ``carries away information about the position of the particle'' \citep[p. 82]{joos:2013} -- that is, induces loss of informational or von Neumann entropy -- but is conservative with respect to energy and momentum. It is a minimal model for position localisation of a quantum particle via the destruction of coherence. More realistic models include noise and dissipation terms but share the central formal feature of \textit{Gaussian-smoothing}. 

Explicitly, the Joos-Zeh master equation takes the form:
\begin{equation}\label{JZequation}
\frac{d\hat{\rho}}{dt} = -\frac{i}{2m} [\hat{p}^2, \hat{\rho}] - \frac{D}{2}[\hat{q}, [\hat{q}, \hat{\rho}]] \end{equation}
where we have assumed a free particle Hamiltonian and the decoherence time scale will be $ t_0= \sqrt{m/D}$. Physically, the localisation rate, $D$, measures how fast interference between different positions disappears for distances smaller than the wavelength of the scattered particles. It has units $\text{cm}^{-2}\:\text{s}^{-1}$ and includes a factor of $\hbar^{-2}$ and a linear dependance on temperature \cite[\S3.2.1]{joos:2013}.

The quantum phase space equation corresponding to \eqref{JZequation} is given by a Fokker-Planck equation for the Wigner function:
\begin{equation}\label{FPeq}
\frac{\partial W}{\partial t}= -\frac{p}{m}\frac{\partial W}{\partial q}+\frac{D}{2}\frac{\partial^2 W}{\partial p^2}
\end{equation}
Although it has the same functional form this equation must not be understood to be equivalent to a Fokker-Planck equation for a classical probability density function since the Wigner function is of course a quasi-probability density and has various non-classical features as per our earlier discussion.   

Following \cite{diosi2002}, the Fokker-Plank equation for the Wigner function can be demonstrated to be equivalent to a progressive Gaussian-smoothing of an initial Wigner function $W(\Gamma;0)$. In particular, we can rewrite the Equation \eqref{FPeq} as a convolution of the form:
\begin{equation}
\label{Wconv}
W(\Gamma;t)=g(\Gamma;\mathbf{C}_W(t))\ast W(x-pt/m,p;0)
\end{equation}
where $g(\Gamma;\mathbf{C}_W(t))$ is a generalised Gaussian function with time-dependent correlation matrix:
\begin{equation}
\mathbf{C}_W(t) = Dt \left(\begin{array}{ccc} t^2/3m^2  & t/2m  &   \\   t/2m & 1  &  \end{array}\right)
\end{equation}
and we have used the $\ast$ symbol for the convolution operation to avoid confusion with the Moyal star product. 

Convolution with a Gaussian function, as per the heat equation, has the general effect of \textit{smoothing} the Wigner function.\footnote{More generally, we can understand decoherence in terms of convolution of the Wigner function with a Gaussian according to a \textit{Weierstrass transform}. This is, in fact, precisely to transform a Wigner function into a Husimi Q-function \cite[\S13]{curtright:2013}. We should not expect the quantum mechanical marginal probabilities to be fully recoverable from the reduced state post-decoherence. Which is perhaps unsurprising.}  The regions of Wigner negativity are of order $\hbar$ and a Gaussian smoothing can be shown to be such that it will progressively render any initial Wigner function positive-definite.\footnote{This is true for Gaussian smoothings but does not hold in general for any averaging \citep{de1990probability}.}  Indeed, \cite{diosi2002} show that by Equation \eqref{Wconv}, \textit{any} initial state will be such that Wigner function will be strictly positive after a finite time $t_D$ which is of the order of the decoherence timescale $ t_0$ defined above. The result of \cite{diosi2002} demonstrates that even for the simplest model of decoherence the dynamical equations smooth out structure of the Wigner function and eliminate Wigner negativity almost immediately.\footnote{See \citep{brody:2024} for a quantum dynamical model that achieves Wigner positivity in finite time without a von Neumann term, i.e. the first commutator on the right hand side of  \eqref{JZequation} in the Joos-Zeh-model, but with a different, but still simple, form of the dissipator term.} Generically, we can expect that Wigner positivity is a novel and robust behaviour that emerges via decoherence based upon the partially interpreted quasi-probability structure as provided by a a quantum possibility space model.

The generic dynamical phenomena of Wigner positivity illustrates the importance of differentiating between classical \textit{probability structures} and classical \textit{possibility space models}. The measure induced by a positive Wigner function are realisations of classical probability structures satisfying the Kolmogorov axioms. However, such a measure \textit{cannot} be interpreted in terms of a classical possibility space models. A positive Wigner function is not equivalent to a classical probability \textit{density function} since the crucial features of local conservation and localisation fail, notwithstanding the Wigner function being positive. On appropriate scales, we will still find the Gaussian-smoothed, positive Wigner function acting in a manner that is irreconcilable with it being a classical probability density and thus,  as it stands, we cannot apply an interpretation in terms of a classical possibility space model. To our knowledge this important point has not previously been highlighted in the literature.  

\subsection{The Role of Semi-Classicality}

The previous sub-section provided a simple illustration of how the non-classical feature of Wigner negativity can be eliminated via decoherence. These methods as a basis for describing the emergence of classicality can generalised to more realistic models. 

\begin{figure}[H]
    \centering
    \centerline{
    \includegraphics[width=0.35\textwidth]{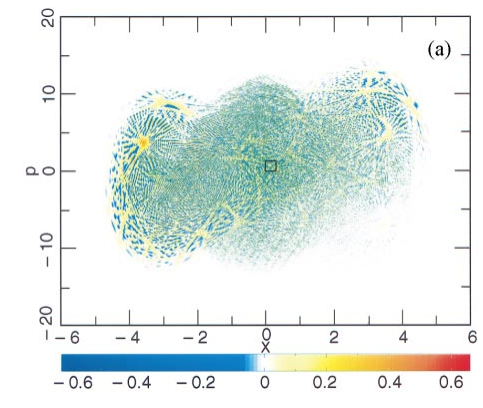}\includegraphics[width=0.35\textwidth]{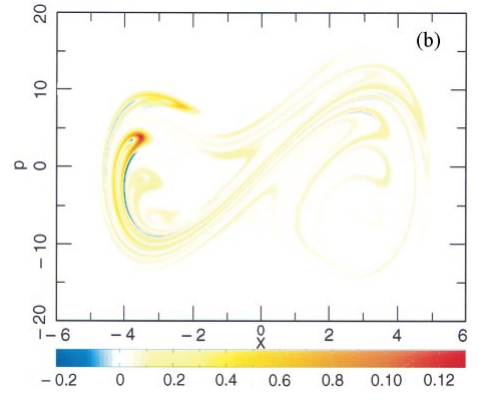}\includegraphics[width=0.35\textwidth]{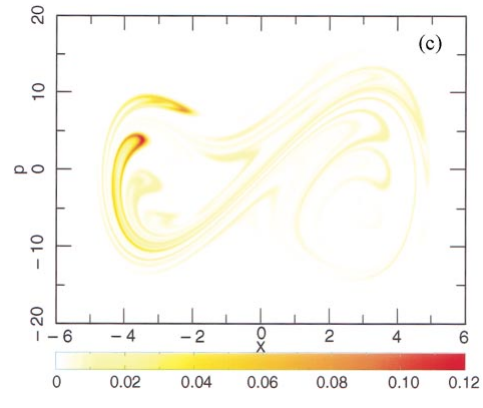}}      \caption{Illustrations of the results of numerical simulations from \cite{Habib:1998} with warm (cold) colours marking regions of positive (negative) density. See text for further explanation. 
    }
    \label{habib}
\end{figure}

Famously, the approach was extended to the study of non-linear models, such as that of the classically chaotic orbit of Hyperion, by \cite{Habib:1998}, leading to the iconic illustrations reproduced in Figure \ref{habib}. The figures (a) and (b) both show the Wigner distribution function from a solution a Fokker-Planck type equation for a non-linear system with a quartic term in the Hamiltonian. The difference between the figures corresponds to solutions to the model at a given time without (a) and with (b) the destruction of large scale quantum coherence. The figure (c) shows the solution of a classical, Fokker-Planck equation for a classical probability density function.  The box represents a phase space area of $4\hbar$. 

The model of \cite{Habib:1998} provides a vivid exemplification of  a quantum possibility space model evolving under an open quantum dynamics that displays a remarkable close correspondence to the behaviour exhibited by a classical possibility space model. \cite{franklin:2023} presents the claim  that we can use appeal to decoherence to frame an account of the emergence classical chaotic phenomenology based upon an underlying \textit{non-chaotic} quantum state in the following terms:\footnote{See \cite{ross:2000,ladyman:2007,Wallace:2010,franklin:2021,mulder:2024}.} 
\begin{quote}
...we may think of the observed classically chaotic orbit of Hyperion as observable evidence of the effects of decoherence in suppressing quantum interference. Classically chaotic Hyperion counts as emergent because much of the structure of the underlying quantum state is conditionally irrelevant to the future dynamics of each classically chaotic Hyperion. In macroscopic terms, what’s screened off are the interference terms that would describe interactions with the Hyperions in other branches -- thus rendering the other branches irrelevant to each branch’s evolution. And the classically chaotic dynamics is not instantiated in the quantum system absent environment induced decoherence. (p. 10)
\end{quote}
There is much to recommend in Franklin's analysis as an account of the relationship between classical and quantum phenomenology. However, one must also bear in mind the foregoing detailed treatment of the structure of decoherence models based upon the Wigner function. A model of decoherence based upon the Wigner function \textit{is} a partial interpretation of a quasi-probability structure via possibility space model. \textit{One is already making use of a partially interpreted generalised probabilistic structure -- a possibility weighting -- when one models decoherence via dynamical equations for the Wigner function as derived from quantum master equations}. 

Purely formal derivations do not have physical content. Classical possibility space models do not emerge \textit{ab initio} from a non-probabilistic and uninterpreted formalism, but rather are emergent in the relevant sense from a partially interpreted, quasi-probabilistic structure. The model of the emergence of classical phenomenology that decoherence models based upon the Wigner function provide is explicitly reliant on its role as a quasi-probability density function in inducing a possibility space weighting. Decoherence must be understood as a basis for quasi-probabilistic emergence rather than non-probabilistic emergence. It is also worth noting that the justification for treating the Wigner function as the preferred representation of quasi-probability relies upon its privileged \textit{empirical} position among quasi-probability representations and corresponding $\star$-product association rules. That is, as noted in the previous sub-section, it is only the Wigner function together with the Moyal $\star$-product that leads to the full set of \textit{experimentally confirmed} marginal distributions corresponding to the Born rule. 

Furthermore, as already noted, Wigner positivity is \textit{necessary but not sufficient} for us to interpret a model as a representation of a classical possibility space due to the failure of localisability and conservation. To understand the relation between the two one is required to provide an account of the \textit{semi-classical limiting relation} between the two which we will in the following section. There is an isomorphism, however, between the dynamical moment equations in quantum and classical possibly space formalisms with the Wigner function and probability density function playing identical roles in the relevant equations, cf. \cite[p.23]{Wallace:2021}. 
    
Consider the dynamical equations for the expectation values (first moments) of position and momentum in the two models. Assume a Hamiltonian of the standard form $H=\frac{p^2}{2m}+V(q)$. Explicit application of the star product as a pseudo-differential operation then gives the expression for the quantum momentum expectation value:
 \begin{eqnarray}
\frac{d \langle p \rangle}{dt} &=&  \langle  \{\{ p,V(q)\}\} \rangle \\
&=&  - \langle  \frac{d V(q)}{dq} \rangle\\
&=& - \int_\Gamma  \frac{d V(q)}{d q} W dq dp
\end{eqnarray}
and for the position expectation value we get:
 \begin{eqnarray}
\frac{d \langle q \rangle}{dt} &=& \frac{1}{2m} \langle  \{\{ q,p^{2}\}\} \rangle \\
&=&  \frac{1}{m} \langle  p \rangle\\
&=& \int_\Gamma  \frac{p}{m} W dq dp
\end{eqnarray}

Following \cite{ballentine:1998}, the corresponding formulas in the classical possibility space model are:
\begin{eqnarray}
\label{meanpeom}
\frac{d \langle p \rangle}{dt} &=& - \int_\Gamma  \frac{d V(q)}{d q} \rho dq dp \\
\frac{d \langle q \rangle}{dt}&=&  \int_\Gamma  \frac{p}{m} \rho dq dp
\end{eqnarray}
We thus have an isomorphism between the classical and quantum probabilistic  phase space formalism. The classical and quantum first moment evolution equations are ``syntactically isomorphic'' \citep{hempel:1965,bartha:2010}. That is, they can be obtained by assigning different physical interpretations to the symbols that appear in a common mathematical form, with the Wigner function in the latter playing the same role as the probability density function in the former. 

\subsection{Emergent Probability}
\label{Emergence}
Our account of the emergence probability in the context of classical and quantum possibility space models will rely upon appeal to a cluster of conceptual innovations from the philosophical literature over the last decades. 

Most foundational is the account of emergence due to  \cite{butterfield:2011}. On Butterfield's account, emergence should be understood to indicate the existence of \textit{novel and robust behaviour} which can be deduced, within a model of that behaviour, by taking an appropriate mathematical limit (this is Butterfield's \textit{1:Deduce}). Furthermore, at least in a weaker (i.e. approximate) sense, such emergence can obtain \textit{on the way to the limit} that forms part of the relevant deduction (this is Butterfield's \textit{2:Before}). Butterfield's account of emergence was subsequently extended in terms of a notion of  coarse-grained emergence by \cite{palacios:2018,palacios:2019,palacios:2022}. We will provide details of that account shortly.

Next, and crucially, our account of the emergence of probability will also require us to draw upon the idea of a `factual' interpretation of the semi-classical $\hbar\rightarrow 0$ limit due to \cite{feintzeig:2020}. According to the account of \cite{feintzeig:2020}, which extends ideas of \cite{rohrlich:1989} and \cite{fletcher:2019}, we should understand the limiting procedure $\hbar\rightarrow 0$ as an \textit{approximation at
certain scales} with the change in the scale determined by the error bounds and units set by a measurement procedure that is used to probe the system. The factual approach allows us to recover the intuitive notion that the limit $\hbar\rightarrow 0$ can be understood in terms of `zooming out' from the characteristic quantum scale set by $\hbar$ by caring less and less about the microscopic details. The contrast is with a `counterfactual' approach to the limiting  $\hbar\rightarrow 0$ procedure in which we consider the physics of possibilities apart from the actual world in which Planck's constant takes a different value that in the limit goes to zero. 

The requirement for the factual approach to the semi-classical limit can be explained most clearly in the context of the contrasting treatment we will adopt towards the $O(\hbar)$ quantities in the theory and the Wigner function. The semi-classical limit of the Wigner function is not always well-behaved \citep{Berry:1977}. Furthermore, for various important classes of quantum states, in particular coherent states, we find that as $\hbar\rightarrow 0$, the Wigner function approximates a $\delta$-function \citep{Berry:1977,curtright:2013,marino:2021}. In such circumstances, applying a `counterfactual' account of the semi-classical limit, where we consider a physical world in which physical dynamics is constituted by the limit $\hbar\rightarrow 0$ behaviour, allows us only to recover the deterministic subset of classical possibility space models since, for consistency, the interpretation requires us to \textit{both} neglect the $O(\hbar)$ terms and work in the limit the probability density is focused on a single point.

By contrast, on the factual approach, since we have different interpretational resources, we can consistently `zoom out' such that terms $O(\hbar)$ are neglected and yet retain non-trivial spread on the Wigner function, even for coherent states. Ultimately, this is to focus on behaviour that, in Butterfield's terms, emerges `on the way to the limit', in the sense that so long as the scale (determined by the error bounds and units set by a measurement procedure) is far enough on the way to the limiting case (in which microscopic details can be neglected entirely), then one is justified in ignoring the $O(\hbar)$ terms precisely because they will be smaller than what is detectable at that scale. This is precisely the intuition one has in considering Figure \ref{habib} and the role of the box of area $4\hbar$. 

Our proposal is then to deploy a combined Butterfield-Palacios-Feintzeig approach to coarse-grained emergence of semi-classical behaviour under a factual interpretation of the semi-classical limit taken as an approximation.  According to \citep[p.39]{palacios:2022}, a coarse-grained description of a system emerges from a fine-grained description, if and only if the former has terms denoting properties or behaviour that are novel and robust with respect to the latter. In our case, the `fine-grained' description is the full quantum phase space model and the `coarse-grained' description is the semi-classical phase space model which is such that the expectation values and expressions truncated $O(\hbar)$ are isomorphic to a classical phase space model.  

We find emergence in the sense of  \cite{palacios:2022} account of coarse-grained emergence specifically since we have: (i) a fine-grained/coarse-grained distinction picked out by phase space areas at order $\hbar$/at order much bigger than $\hbar$; 
(ii) the coarse-grained description has features that are not features of the fine-grained description, specifically local conservation and localisation of the (quasi)-probability density; (iii) the behaviour represented by the fine-grained description exists at the same time as the behaviour represented by the coarse-grained description (i.e. we have \textit{synchronic emergence}); (iv) the coarse-grained description refers to behaviours (local conservation and localisation) that are insensitive to (robust under) variation of the microphysical details that characterise a particular token (for example, with respect to variation of  fine-grained details $O(\hbar^2)$); (v) the coarse-grained level depends on the fine-grained level in the sense that every change in the coarse-grained level must imply a change in the fine-grained level (i.e. we have supervenience).

Quasi-probabilistic emergence via decoherence consists in the derivation of the novel and robust behaviour of Wigner positivity at finite values of the relevant parameters (time and temperature in realistic models). Coarse-grained emergence via a factual interpretation of the semi-classical limit consists in the derivation of the behaviours of conservation and localisation in the coarse-grained description that are novel and robust with respect to the fine-grained description.  The quasi-probabilistic emergence via decoherence is dependant on the decoherence time scale and is thus `diachronic'. By contrast, the semi-classical emergence is dependent upon phase space areas in units of $\hbar$ and is thus `synchronic'. A schematic diagram for the relevant pattern of interrelations is provided in Figure \ref{boxes}.  The combined double emergence picture is sufficient to deduce a classical possibility space model from a quantum possibility space model. Classical probability in phase space is emergent from an underlying quantum phase space dynamics.

We take ourselves to have provided a comprehensive formal and physical account of the emergence of probability from quantum theory. This is not, however, an interpretation of the theory or a solution to the measurement problem.  Rather, our aim has been to clarify the sense in which principles of quantum mechanics require generalised probabilistic concepts and how such concepts and how such generalised can be connected to classical probability via physical arguments. In the following section, we will summarise our main arguments and offer an outlook regarding the further work that would be required to offer a response to the interpretational issue.

\begin{figure}[H]

\begin{center}
\begin{tikzpicture}

% Define nodes with explicit coordinates
\node[draw, rectangle, thick, minimum width=2.5cm, minimum height=2cm, inner sep=6pt] (Model_R) at (6,2) {Classical PSM};
\node[draw, rectangle, thick, minimum width=2.5cm, minimum height=2cm, inner sep=6pt] (Model_I) at (6,-4) {Wigner-Positive PSM};
\node[draw, rectangle, thick, minimum width=2.5cm, minimum height=2cm, inner sep=6pt] (Target) at (-2,-4) {Quantum PSM};

\draw[->, thick, >=stealth] (Model_I) -- node[midway, left] {\parbox{4cm}{\centering \textit{Semi-classical emergence (synchronic)}}} (Model_R);
\draw[->, thick, >=stealth] (Target) -- node[midway, above] {\parbox{4cm}{\centering \textit{Quasi-probabilistic emergence (diachronic)}}} (Model_I);

\end{tikzpicture}
\end{center}
\caption{Schematic diagram showing relationship between Quantum and Classical Possibility Space Models (PSMs). Inspired by \cite[Fig.9]{palacios:2022}.}
\label{boxes}
\end{figure}

\section{Recapitulation and Outlook}
\label{Outlook}
Let us return to the original dialectic with which we started our analysis of probability  and decoherence. Recall, in particular, that the arguments of \cite{dawid:2015} were that a certain package of interpretative moves concerning probability and the quantum formalism leads to an incoherent conclusion. The foregoing analysis allows us to consider the contraposition this argument. That is, we have sought to establish a framework for the analysis of classical and quantum probability within which any coherent interpretation must be expected to operate. 

On our analysis, an account of the role of probability in quantum mechanics can most plausibly play out in only one of two ways. First, probability can be introduced as a fully formed classical probability in connection with an extra posit such as collapse, hidden variables, or observers. Second, one can abstain from extra posits, and establish the probabilistic nature of quantum mechanics as an approximate, emergent concept. In the latter case, there is no plausible way to avoid adding to pure wave mechanics a partial interpretation in terms of possibility weightings. In particular,  there is no way to understand decoherence in general, or the suppression of small amplitudes in particular, absent a partially interpreted structure that weights possibilities. The requirement for such a partial interpretation does not render the Many Worlds interpretation incoherent in itself. It does, however, place strong constraints upon the way in which such an interpretation can be packaged together with an approach to probability and possibility. In particular, it shows that there is no coherent prospect for an interpretational package that seeks to combine an entirely non-probabilistic account of the emergence of `words' with a post-decoherence decision theoretic derivation of probability.  In this sense the claims of \cite{dawid:2015} can be understood to be vindicated against those of \cite{Saunders:2021} and \cite{franklin:2023}. 

More importantly, our analysis indicates that any \textit{full interpretation} of quantum mechanics that does not seek to introduce probability via extra posits must grapple with the quasi-probabilistic nature of the theory. That is, if probability is not introduced as a fully formed classical concept in connection with an extra posit such as collapse, hidden variables, measurement, or observers, then we will need to find a way to attribute physical significance to quasi-probabilities (or quasi-measures) at the level of the fundamental theory. Arguments from similarity, do not provide a solution to this problem. As a conceptual basis for neglecting small amplitudes they fail; and using them as merely heuristic reasons for adding to quantum mechanics a prescription to neglect small amplitudes would subvert precisely the most attractive feature of Many Worlds interpretations: that of requiring no posits beyond the wave function equations. 

Let us conclude by briefly outlining a selection of issues and ideas relevant to the extension of our project towards a full interpretation. First, it would be interesting to consider the connection between our account of the emergence of probability and work of \cite{feintzeig:2017}. These results draw connections between  non-contextual hidden variable interpretations and the existence of a finite null cover and this would appear to make difficult certain attempts to move from a partial to full interpretation of the quasi-measure over possibility space. We can thus appeal to such result to place constraints of the form that a full interpretation of generalised probability within the theory could take.  

Second, there is an interesting connection between our analysis, proposals for quantum measure theory \citep{sorkin:2010,clements:2017}, and generalised probabilistic  structures found within the decoherent histories programme  \citep{gellman:1996,halliwell:2010}. In particular, it can be proved that the diagonal elements of the decoherence functional are equivalent to a `quantal-measure' which is a specific form of our quasi-measure that obeys a particular (non-classical) sum rule on the algebra of events \citep{sorkin:1994,dowker:2022}. The decoherent histories framework thus \textit{is} a partial interpretation of a quasi-probability structure in precisely our terms since it involves the weighting of quantum possibilities via the decoherence functional.  Since there is an explicit dependence on coarse-graining in this approach to the emergence of classical probability there is a plausible path for both reconstructing our analysis in histories terms and applying such a quantum measure theory version of the programme as a full interpretation of the theory.

Third, it would be interesting to consider the connection between our analysis and the recent work on probability in the Everett interpretation due to \cite{saunders:2024}. This work includes a proposal for a `finite frequentist' approach to quantum probability within the many world interpretation that plausibly fulfils to requirements that we have argued for. In particular, it appears to provide an interpretation of \textit{possibility weightings as finite frequencies} within the full quantum context without appeal to decoherence. Interestingly, the notion of quantum probability identified by Saunders is a non-Kolmogorovian \textit{imprecise probability}, as provide by interval probabilities that do not satisfy the additivity requirement. It would be an interesting formal and philosophical project to recast Saunders approach within the quantum phase space formalism to allow for direct comparison with our own work.    

Fourth, much more could be said about the interaction between diachronic and synchronic limits in the context of classical and quantum probability. For example, it would be of significant physical and philosophical interest to more fully understand the relation between the semi-classical $\hbar\rightarrow 0$ and long time $t\rightarrow  \infty$ limits. It was recently demonstrated by  \cite{bonds:2024}, that application of the method of arbitrary functions allows one to explicitly derive a classical possibility model in the case of the quantum oscillator via the combined $t\rightarrow \infty$ and $\hbar\rightarrow 0$ limits. Furthermore, this approach allows for the recovery of the Born rule when combined with a toy model of quantum measurement in terms of a perturbation treated as a random variable with (almost) arbitrary initial probability distribution in a manner inspired by the `flea' perturbation of \cite{landsman:2013}. This approach would offer a further alternative strategy for extending our analysis towards a full interpretation of the theory.\footnote{Relatedly, it would  also be interesting to consider the relation to the work of \cite{layton:2023} and \cite{hernandez:2023}.}

\section*{Acknowledgements} 

Work on this paper has drawn on a large number of discussions with a wide variety of helpful colleagues over the last few years. We are particularly grateful to Simon Saunders and Alex Franklin for continued highly enjoyable interactions and to Ben Feintzeig for constructive critical engagement. Further thanks are also due to Jim Al-Khalili, Kevin Blackwell, Dorje Brody, Adam Caulton, Fay Dowker, Sam Fletcher, Gui Franzmann, Simon Friedrich, Sean Gryb, Jonte Hance, Nick Huggett, Eddy Keming Chen, Jason Konek, James Ladyman, Tushar Menon, Richard Pettigrew, James Read, Christian R\"oken, Tony Short, Jer Steeger, Mritunjay Tyagi, Cristi Stoica, and Lev Vaidman, and three anonymous referees (apologies if we have missed anyone out). Thanks also to audiences in Surrey, Bonn, Oxford, and Bristol. We are also appreciative to Cosmas Zachos for helpful written comments on a draft manuscript. Work on this paper was supported by the John Templeton Foundation grant number 62210. The views expressed may not represent the opinions of the Foundation. This research was in part funded by the Swedish Research Council grant number 2022-01893\_VR.


\begin{thebibliography}{}

\bibitem[\protect\citeauthoryear{Adlam}{Adlam}{2014}]{Adlam:2014}
Adlam, E. (2014, 8).
\newblock The problem of confirmation in the Everett interpretation.
\newblock {\em Studies in History and Philosophy of Science Part B: Studies in
  History and Philosophy of Modern Physics\/}~{\em 47\/}(0), 21--32.
  \newblock \href{http://dx.doi.org/10.1016/j.shpsb.2014.03.004}{DOI:10.1016/j.shpsb.2014.03.004}

\bibitem[\protect\citeauthoryear{Andreas}{Andreas}{2021}]{sep-theoretical-terms-science}
Andreas, H. (2021).
\newblock {Theoretical Terms in Science}.
\newblock In E.~N. Zalta (Ed.), {\em The {Stanford} Encyclopedia of
  Philosophy\/} ({F}all 2021 ed.). Metaphysics Research Lab, Stanford
  University.
  \newblock \href{https://plato.stanford.edu/archives/fall2021/entries/theoretical-terms-science/}{URL:plato.stanford.edu/archives/fall2021/entries/theoretical-terms-science/}

\bibitem[\protect\citeauthoryear{Aniello}{Aniello}{2016}]{aniello:2016}
Aniello, P. (2016).
\newblock Functions of positive type on phase space, between classical and
  quantum, and beyond.
\newblock In {\em Journal of Physics: Conference Series}, Volume 670, pp.\
  012004. IOP Publishing.
  \newblock \href{https://iopscience.iop.org/article/10.1088/1742-6596/670/1/012004}{DOI: 10.1088/1742-6596/670/1/012004}

\bibitem[\protect\citeauthoryear{Arageorgis, Earman, and Ruetsche}{Arageorgis
  et~al.}{2017}]{Arageorgis:2017}
Arageorgis, A., J.~Earman, and L.~Ruetsche (2017).
\newblock Additivity requirements in classical and quantum probability.
\newblock \href{https://philsci-archive.pitt.edu/id/eprint/13024}{URL: https://philsci-archive.pitt.edu/id/eprint/13024}

\bibitem[\protect\citeauthoryear{Baker}{Baker}{2007}]{Baker:2007}
Baker, D.~J. (2007).
\newblock Measurement outcomes and probability in Everettian quantum mechanics.
\newblock {\em Studies In History and Philosophy of Science Part B: Studies In
  History and Philosophy of Modern Physics\/}~{\em 38\/}(1), 153 -- 169.
  \newblock \href{https://doi.org/10.1016/j.shpsb.2006.05.003}{DOI: 10.1016/j.shpsb.2006.05.003}

\bibitem[\protect\citeauthoryear{Ballentine and McRae}{Ballentine and
  McRae}{1998}]{ballentine:1998}
Ballentine, L. and S.~McRae (1998).
\newblock Moment equations for probability distributions in classical and
  quantum mechanics.
\newblock {\em Physical Review A\/}~{\em 58\/}(3), 1799.
\newblock \href{https://doi.org/10.1103/PhysRevA.58.1799}{DOI: 10.1103/PhysRevA.58.1799}

\bibitem[\protect\citeauthoryear{Bartha}{Bartha}{2010}]{bartha:2010}
Bartha, P. (2010).
\newblock {\em By parallel reasoning}.
\newblock Oxford University Press.
\newblock \href{https://doi.org/10.1093/acprof:oso/9780195325539.001.0001}{DOI: 10.1093/acprof:oso/9780195325539.001.0001}

\bibitem[\protect\citeauthoryear{Berry}{Berry}{1977}]{Berry:1977}
Berry, M.~V. (1977).
\newblock Semi-classical mechanics in phase space: a study of Wigner's
  function.
\newblock {\em Philosophical Transactions of the Royal Society of London.
  Series A, Mathematical and Physical Sciences\/}~{\em 287\/}(1343), 237--271.
  \newblock \href{https://doi.org/10.1098/rsta.1977.0145}{DOI: 10.1098/rsta.1977.0145}

\bibitem[\protect\citeauthoryear{Bonds, Burson, Cicchella, Feintzeig, Yusaini,
  et~al.}{Bonds et~al.}{2024}]{bonds:2024}
Bonds, L., B.~Burson, K.~Cicchella, B.~H. Feintzeig, A.~Yusaini, et~al. (2024).
\newblock Quantum probability via the method of arbitrary functions.
\newblock \href{ 	
https://doi.org/10.48550/arXiv.2409.16457}{DOI:10.48550/arXiv.2409.16457}

\bibitem[\protect\citeauthoryear{Booth, Chabaud, and Emeriau}{Booth
  et~al.}{2022}]{booth:2022}
Booth, R.~I., U.~Chabaud, and P.-E. Emeriau (2022).
\newblock Contextuality and Wigner negativity are equivalent for
  continuous-variable quantum measurements.
\newblock {\em Physical Review Letters\/}~{\em 129\/}(23), 230401.
 \newblock \href{https://doi.org/10.1103/PhysRevLett.129.230401}{DOI: 10.1103/PhysRevLett.129.230401}

\bibitem[\protect\citeauthoryear{Brody, Graefe, and Melanathuru}{Brody
  et~al.}{2025}]{brody:2024}
Brody, D.~C., E.-M. Graefe, and R.~Melanathuru (2025).
\newblock Phase-space measurements, decoherence and classicality.
\newblock {\em Physical Review Letters\/}~{\em 134\/}(12), 120201.
 \newblock \href{https://doi.org/10.1103/PhysRevLett.134.120201}{DOI: https://doi.org/10.1103/PhysRevLett.134.120201}

\bibitem[\protect\citeauthoryear{Brown and Porath}{Brown and
  Porath}{2020}]{brown:2020}
Brown, H.~R. and G.~B. Porath (2020).
\newblock Everettian probabilities, the Deutsch-Wallace theorem and the
  principal principle.
\newblock {\em Quantum, probability, logic: the work and influence of Itamar
  Pitowsky\/}, 165--198.
   \newblock \href{https://doi.org/10.1007/978-3-030-34316-3}{DOI: 10.1007/978-3-030-34316-3}

\bibitem[\protect\citeauthoryear{Butterfield}{Butterfield}{2011}]{butterfield:2011}
Butterfield, J. (2011).
\newblock Less is different: emergence and reduction reconciled.
\newblock {\em Foundations of Physics\/}~{\em 41\/}(6), 1065--1135.
 \newblock \href{https://doi.org/10.1007/s10701-010-9516-1}{DOI: 10.1007/s10701-010-9516-1}

\bibitem[\protect\citeauthoryear{Carnap}{Carnap}{1958}]{carnap:1958}
Carnap, v.~R. (1958).
\newblock Beobachtungssprache und Theoretische Sprache.
\newblock {\em Dialectica\/}~{\em 12\/}(3-4), 236--248.
 \newblock \href{https://doi.org/10.1111/j.1746-8361.1958.tb01461.x}{DOI: 10.1111/j.1746-8361.1958.tb01461.x}

\bibitem[\protect\citeauthoryear{Case}{Case}{2008}]{case:2008}
Case, W.~B. (2008).
\newblock Wigner functions and Weyl transforms for pedestrians.
\newblock {\em American Journal of Physics\/}~{\em 76\/}(10), 937--946.
 \newblock \href{
https://doi.org/10.1119/1.2957889
}{
DOI: 10.1119/1.2957889
}

\bibitem[\protect\citeauthoryear{Clements, Dowker, and Wallden}{Clements
  et~al.}{2017}]{clements:2017}
Clements, K., F.~Dowker, and P.~Wallden (2017).
\newblock Physical logic.
\newblock {\em The Incomputable: Journeys Beyond the Turing Barrier\/}, 47--61.
 \newblock \href{https://doi.org/10.1007/978-3-319-43669-2}{DOI: 10.1007/978-3-319-43669-2}

\bibitem[\protect\citeauthoryear{Cohen}{Cohen}{1966}]{cohen:1966}
Cohen, L. (1966).
\newblock Can quantum mechanics be formulated as a classical probability
  theory?
\newblock {\em Philosophy of Science\/}~{\em 33\/}(4), 317--322.
 \newblock \href{https://doi.org/10.1086/288104}{DOI:  10.1086/288104 }

\bibitem[\protect\citeauthoryear{Curtright, Fairlie, and Zachos}{Curtright
  et~al.}{2013}]{curtright:2013}
Curtright, T.~L., D.~B. Fairlie, and C.~K. Zachos (2013).
\newblock {\em A concise treatise on quantum mechanics in phase space}.
\newblock World Scientific Publishing Company.
 \newblock \href{https://doi.org/10.1142/8870 }{DOI: 10.1142/8870 }

\bibitem[\protect\citeauthoryear{Dahl, Mack, Wolf, and Schleich}{Dahl
  et~al.}{2006}]{dahl:2006}
Dahl, J.~P., H.~Mack, A.~Wolf, and W.~P. Schleich (2006).
\newblock Entanglement versus negative domains of Wigner functions.
\newblock {\em Physical Review A---Atomic, Molecular, and Optical
  Physics\/}~{\em 74\/}(4), 042323.
   \newblock \href{https://doi.org/10.1103/PhysRevA.74.042323}{DOI: 10.1103/PhysRevA.74.042323}

\bibitem[\protect\citeauthoryear{Dawid and Th{\'e}bault}{Dawid and
  Th{\'e}bault}{2014}]{Dawid:2014}
Dawid, R. and K.~P. Th{\'e}bault (2014).
\newblock Against the empirical viability of the Deutsch--Wallace--Everett
  approach to quantum mechanics.
\newblock {\em Studies in History and Philosophy of Science Part B: Studies in
  History and Philosophy of Modern Physics\/}~{\em 47\/}(0), 55 -- 61.
   \newblock \href{https://doi.org/10.1016/j.shpsb.2014.05.005}{DOI: 10.1016/j.shpsb.2014.05.005}

\bibitem[\protect\citeauthoryear{Dawid and Th{\'e}bault}{Dawid and
  Th{\'e}bault}{2015}]{dawid:2015}
Dawid, R. and K.~P. Th{\'e}bault (2015).
\newblock Many worlds: decoherent or incoherent?
\newblock {\em Synthese\/}~{\em 192}, 1559--1580.
 \newblock \href{https://doi.org/10.1007/s11229-014-0650-8}{DOI: 10.1007/s11229-014-0650-8}

\bibitem[\protect\citeauthoryear{de~Aguiar and de~Almeida}{de~Aguiar and
  de~Almeida}{1990}]{de1990probability}
de~Aguiar, M.~A. and A.~O. de~Almeida (1990).
\newblock On the probability density interpretation of smoothed Wigner
  functions.
\newblock {\em Journal of Physics A: Mathematical and General\/}~{\em
  23\/}(19), L1025.
   \newblock \href{https://doi.org/10.1088/0305-4470/23/19/002}{DOI: 10.1088/0305-4470/23/19/002}

\bibitem[\protect\citeauthoryear{De~Gosson}{De~Gosson}{2017}]{Gosson:2017}
De~Gosson, M.~A. (2017).
\newblock {\em The Wigner Transform}.
\newblock World Scientific Publishing Company.
 \newblock \href{https://doi.org/10.1142/q0089}{DOI: 10.1142/q0089}

\bibitem[\protect\citeauthoryear{Delfosse, Okay, Bermejo-Vega, Browne, and
  Raussendorf}{Delfosse et~al.}{2017}]{delfosse:2017}
Delfosse, N., C.~Okay, J.~Bermejo-Vega, D.~E. Browne, and R.~Raussendorf
  (2017).
\newblock Equivalence between contextuality and negativity of the Wigner
  function for qudits.
\newblock {\em New Journal of Physics\/}~{\em 19\/}(12), 123024.
 \newblock \href{https://doi.org/10.1088/1367-2630/aa8fe3}{DOI: 10.1088/1367-2630/aa8fe3}

\bibitem[\protect\citeauthoryear{Dell'Antonio}{Dell'Antonio}{2016}]{delllectures}
Dell'Antonio, G. (2016).
\newblock {\em Lectures on the Mathematics of Quantum Mechanics II: Selected
  Topics}.
\newblock Springer.
 \newblock \href{https://doi.org/10.2991/978-94-6239-115-4}{DOI: 10.2991/978-94-6239-115-4}

\bibitem[\protect\citeauthoryear{Deutsch}{Deutsch}{1999}]{Deutsch:1999}
Deutsch, D. (1999).
\newblock Quantum theory of probability and decisions.
\newblock {\em Proceedings of the Royal Society of London. Series A:
  Mathematical, Physical and Engineering Sciences\/}~{\em 455\/}(1988),
  3129--3137.
   \newblock \href{https://doi.org/10.1098/rspa.1999.0443}{DOI: 10.1098/rspa.1999.0443}

\bibitem[\protect\citeauthoryear{Dias, de~Gosson, and Prata}{Dias
  et~al.}{2019}]{dias:2019}
Dias, N.~C., M.~A. de~Gosson, and J.~N. Prata (2019).
\newblock A refinement of the Robertson--Schr{\"o}dinger uncertainty principle
  and a Hirschman--Shannon inequality for Wigner distributions.
\newblock {\em Journal of Fourier Analysis and Applications\/}~{\em 25\/}(1),
  210--241.
   \newblock \href{https://doi.org/10.1007/s00041-018-9602-x}{DOI: 10.1007/s00041-018-9602-x}

\bibitem[\protect\citeauthoryear{Di{\'o}si and Kiefer}{Di{\'o}si and
  Kiefer}{2002}]{diosi2002}
Di{\'o}si, L. and C.~Kiefer (2002).
\newblock Exact positivity of the Wigner and p-functions of a Markovian open
  system.
\newblock {\em Journal of Physics A: Mathematical and General\/}~{\em
  35\/}(11), 2675.
   \newblock \href{https://doi.org/10.1088/0305-4470/35/11/312}{DOI: 10.1088/0305-4470/35/11/312}

\bibitem[\protect\citeauthoryear{Dizadji-Bahmani}{Dizadji-Bahmani}{2013}]{dizadji:2013}
Dizadji-Bahmani, F. (2013).
\newblock {The Probability Problem in Everettian Quantum Mechanics Persists}.
\newblock {\em The British Journal for the Philosophy of Science\/}, axt035.
 \newblock \href{https://doi.org/10.1093/bjps/axt035}{DOI: 10.1093/bjps/axt035}

\bibitem[\protect\citeauthoryear{Dowker and Wilkes}{Dowker and
  Wilkes}{2022}]{dowker:2022}
Dowker, F. and H.~Wilkes (2022).
\newblock An argument for strong positivity of the decoherence functional in
  the path integral approach to the foundations of quantum theory.
\newblock {\em AVS Quantum Science\/}~{\em 4\/}(1).
 \newblock \href{
https://doi.org/10.1116/5.0073587
}{DOI: 10.1116/5.0073587
}

\bibitem[\protect\citeauthoryear{Feintzeig}{Feintzeig}{2020}]{feintzeig:2020}
Feintzeig, B.~H. (2020).
\newblock The classical limit as an approximation.
\newblock {\em Philosophy of Science\/}~{\em 87\/}(4), 612--639.
 \newblock \href{https://doi.org/10.1086/709731}{DOI: 10.1086/709731}

\bibitem[\protect\citeauthoryear{Feintzeig and Fletcher}{Feintzeig and
  Fletcher}{2017}]{feintzeig:2017}
Feintzeig, B.~H. and S.~C. Fletcher (2017).
\newblock On noncontextual, non-Kolmogorovian hidden variable theories.
\newblock {\em Foundations of Physics\/}~{\em 47}, 294--315.
 \newblock \href{https://doi.org/10.1007/s10701-017-0061-z}{DOI: 10.1007/s10701-017-0061-z}

\bibitem[\protect\citeauthoryear{Feller}{Feller}{1991}]{feller:1991}
Feller, W. (1991).
\newblock {\em An introduction to probability theory and its applications,
  Volume 2}, Volume~81.
\newblock John Wiley \& Sons.

\bibitem[\protect\citeauthoryear{Fine}{Fine}{1982a}]{fine:1982}
Fine, A. (1982a).
\newblock Hidden variables, joint probability, and the Bell inequalities.
\newblock {\em Physical Review Letters\/}~{\em 48\/}(5), 291.
 \newblock \href{https://doi.org/10.1103/PhysRevLett.48.291}{DOI: 10.1103/PhysRevLett.48.291}

\bibitem[\protect\citeauthoryear{Fine}{Fine}{1982b}]{fine:1982b}
Fine, A. (1982b).
\newblock Joint distributions, quantum correlations, and commuting observables.
\newblock {\em Journal of Mathematical Physics\/}~{\em 23\/}(7), 1306--1310.
 \newblock \href{https://doi.org/10.1063/1.525514
}{DOI: 10.1063/1.525514}

\bibitem[\protect\citeauthoryear{Fletcher}{Fletcher}{2019}]{fletcher:2019}
Fletcher, S.~C. (2019).
\newblock On the reduction of general relativity to Newtonian gravitation.
\newblock {\em Studies in History and Philosophy of Science Part B: Studies in
  History and Philosophy of Modern Physics\/}~{\em 68}, 1--15.
   \newblock \href{https://doi.org/10.1016/j.shpsb.2019.04.005}{DOI: 10.1016/j.shpsb.2019.04.005}

\bibitem[\protect\citeauthoryear{Franklin}{Franklin}{2023}]{franklin:2023}
Franklin, A. (2023).
\newblock Incoherent? no, just decoherent: How quantum many worlds emerge.
\newblock {\em Philosophy of Science\/}~{\em 91} (2),  288--309.
\newblock \href{https://doi.org/10.1017/psa.2023.155 }{DOI: 10.1017/psa.2023.155 }


\bibitem[\protect\citeauthoryear{Franklin and Robertson}{Franklin and
  Robertson}{2021}]{franklin:2021}
Franklin, A. and K.~Robertson (2021).
\newblock Emerging into the rainforest: Emergence and special science ontology.
\newblock {\em European Journal for Philosophy of Science\/}~{\em 14\/}(61).
 \newblock \href{https://doi.org/10.1007/s13194-024-00622-4}{DOI: 10.1007/s13194-024-00622-4}

\bibitem[\protect\citeauthoryear{Friederich}{Friederich}{2021}]{friederich:2021}
Friederich, S. (2021).
\newblock Introducing the Q-based interpretation of quantum theory.
\newblock {\em British Journal for the Philosophy of Science\/}~{\em
  doi.org.10.1086/716196}.
   \newblock \href{https://doi.org/10.1086/716196}{DOI: 10.1086/716196}

\bibitem[\protect\citeauthoryear{Gell-Mann and Hartle}{Gell-Mann and
  Hartle}{1996}]{gellman:1996}
Gell-Mann, M. and J.~B. Hartle (1996).
\newblock Quantum mechanics in the light of quantum cosmology.
\newblock In {\em Foundations Of Quantum Mechanics In The Light Of New
  Technology: Selected Papers from the Proceedings of the First through Fourth
  International Symposia on Foundations of Quantum Mechanics}, pp.\  347--369.
  World Scientific.
   \newblock DOI: 10.1142/9789812819895 0036

\bibitem[\protect\citeauthoryear{Gibbs}{Gibbs}{1902}]{gibbs:1902}
Gibbs, J.~W. (1902).
\newblock {\em Elementary principles in statistical mechanics: developed with
  especial reference to the rational foundations of thermodynamics}.
\newblock C. Scribner's sons.
 \newblock \href{https://doi.org/10.1017/CBO9780511686948}{DOI: 10.1017/CBO9780511686948}

\bibitem[\protect\citeauthoryear{Greaves and Myrvold}{Greaves and
  Myrvold}{2010}]{Greaves:2010}
Greaves, H. and W.~Myrvold (2010).
\newblock Everett and evidence.
\newblock In S.~Saunders, J.~Barrett, A.~Kent, and D.~Wallace (Eds.), {\em Many
  Worlds? Everett, Quantum Theory, and Reality}, Chapter~9, pp.\  264--304.
  Oxford Univeristy Press.
   \newblock \href{https://doi.org/10.1093/acprof:oso/9780199560561.001.0001}{DOI: 10.1093/acprof:oso/9780199560561.001.0001}

\bibitem[\protect\citeauthoryear{Habib, Shizume, and Zurek}{Habib
  et~al.}{1998}]{Habib:1998}
Habib, S., K.~Shizume, and W.~H. Zurek (1998).
\newblock {Decoherence, chaos, and the correspondence principle}.
\newblock {\em Phys. Rev. Lett.\/}~{\em 80}, 4361--4365.
 \newblock \href{https://doi.org/10.1103/PhysRevLett.80.4361}{DOI: 10.1103/PhysRevLett.80.4361}

\bibitem[\protect\citeauthoryear{Halliwell}{Halliwell}{2010}]{halliwell:2010}
Halliwell, J. (2010).
\newblock Macroscopic superpositions, decoherent histories, and the emergence
  of hydrodynamic behaviour.
  \newblock In S.~Saunders, J.~Barrett, A.~Kent, and D.~Wallace (Eds.), {\em Many
  Worlds? Everett, Quantum Theory, and Reality}, pp.\  99--117..
  Oxford Univeristy Press.
   \newblock \href{https://doi.org/10.1093/acprof:oso/9780199560561.001.0001}{DOI: 10.1093/acprof:oso/9780199560561.001.0001}


\bibitem[\protect\citeauthoryear{Hartmann}{Hartmann}{2015}]{Hartmann:2015}
Hartmann, S. (2015).
\newblock Imprecise probabilities in quantum mechanics.
\newblock In C.~E. Crangle, A.~G. de~la Sienra, and H.~E. Longino (Eds.), {\em
  Foundations and Methods From Mathematics to Neuroscience: Essays Inspired by
  Patrick Suppes}, pp.\  77--82. Stanford Univ Center for the Study.

\bibitem[\protect\citeauthoryear{Healey}{Healey}{2017}]{healey:2017}
Healey, R. (2017).
\newblock {\em The quantum revolution in philosophy}.
\newblock Oxford University Press.
 \newblock  \href{https://doi.org/10.1093/oso/9780198714057.001.0001}{DOI: 10.1093/oso/9780198714057.001.0001}

\bibitem[\protect\citeauthoryear{Hempel}{Hempel}{1965}]{hempel:1965}
Hempel, C.~G. (1965).
\newblock {\em Aspects of Scientific Explanation and Other Essays in the Philosophy of Science}.
\newblock New York: Free Press.

\bibitem[\protect\citeauthoryear{Hern{\'a}ndez, Ranard, and
  Riedel}{Hern{\'a}ndez et~al.}{2023}]{hernandez:2023}
Hern{\'a}ndez, F., D.~Ranard, and C.~J. Riedel (2023).
\newblock The $\hbar$ to $0$ limit of open quantum systems with general
  lindbladians: vanishing noise ensures classicality beyond the ehrenfest time.
\newblock {\em arXiv preprint arXiv:2307.05326\/}.
 \newblock \href{ 	
https://doi.org/10.48550/arXiv.2307.05326}{ 	
DOI: 10.48550/arXiv.2307.05326}

\bibitem[\protect\citeauthoryear{Hillery, O'Connell, Scully, and
  Wigner}{Hillery et~al.}{1984}]{hillery:1984}
Hillery, M., R.~F. O'Connell, M.~O. Scully, and E.~P. Wigner (1984).
\newblock Distribution functions in physics: Fundamentals.
\newblock {\em Physics reports\/}~{\em 106\/}(3), 121--167.
 \newblock \href{https://doi.org/10.1016/0370-1573(84)90160-1}{DOI: 10.1016/0370-1573(84)90160-1}

\bibitem[\protect\citeauthoryear{Hudson}{Hudson}{1974}]{hudson:1974}
Hudson, R.~L. (1974).
\newblock When is the Wigner quasi-probability density non-negative?
\newblock {\em Reports on Mathematical Physics\/}~{\em 6\/}(2), 249--252.
 \newblock \href{https://doi.org/10.1016/0034-4877(74)90007-X}{DOI: 10.1016/0034-4877(74)90007-X}

\bibitem[\protect\citeauthoryear{Huggett, Lizzi, and Menon}{Huggett
  et~al.}{2021}]{huggett:2021}
Huggett, N., F.~Lizzi, and T.~Menon (2021).
\newblock Missing the point in noncommutative geometry.
\newblock {\em Synthese\/}, 1--34.
 \newblock \href{https://doi.org/10.1007/s11229-020-02998-1}{DOI: 10.1007/s11229-020-02998-1}

\bibitem[\protect\citeauthoryear{Joos and Zeh}{Joos and Zeh}{1985}]{joos:1985}
Joos, E. and H.~D. Zeh (1985).
\newblock The emergence of classical properties through interaction with the
  environment.
\newblock {\em Zeitschrift f{\"u}r Physik B Condensed Matter\/}~{\em 59},
  223--243.
   \newblock \href{https://doi.org/10.1007/BF01725541}{DOI: 10.1007/BF01725541}

\bibitem[\protect\citeauthoryear{Joos, Zeh, Kiefer, Giulini, Kupsch, and
  Stamatescu}{Joos et~al.}{2013}]{joos:2013}
Joos, E., H.~D. Zeh, C.~Kiefer, D.~J. Giulini, J.~Kupsch, and I.-O. Stamatescu
  (2013).
\newblock {\em Decoherence and the appearance of a classical world in quantum
  theory}.
\newblock Springer Science \& Business Media.
 \newblock \href{https://doi.org/10.1007/978-3-662-05328-7}{DOI: 10.1007/978-3-662-05328-7}

\bibitem[\protect\citeauthoryear{Ladyman and Ross}{Ladyman and
  Ross}{2007}]{ladyman:2007}
Ladyman, J. and D.~Ross (2007).
\newblock {\em Every thing must go: Metaphysics naturalized}.
\newblock Oxford University Press.
 \newblock \href{https://doi.org/10.1093/acprof:oso/9780199276196.001.0001}{DOI: 10.1093/acprof:oso/9780199276196.001.0001}

\bibitem[\protect\citeauthoryear{Landsman}{Landsman}{2012}]{landsman:2012}
Landsman, N.~P. (2012).
\newblock {\em Mathematical topics between classical and quantum mechanics}.
\newblock Springer Science \& Business Media.
 \newblock \href{https://doi.org/10.1007/978-1-4612-1680-3}{DOI: 10.1007/978-1-4612-1680-3}

\bibitem[\protect\citeauthoryear{Landsman and Reuvers}{Landsman and
  Reuvers}{2013}]{landsman:2013}
Landsman, N.~P. and R.~Reuvers (2013).
\newblock A flea on schr{\"o}dinger's cat.
\newblock {\em Foundations of Physics\/}~{\em 43}, 373--407.
 \newblock \href{https://doi.org/10.1007/s10701-013-9700-1}{DOI: 10.1007/s10701-013-9700-1}

\bibitem[\protect\citeauthoryear{Layton and Oppenheim}{Layton and
  Oppenheim}{2023}]{layton:2023}
Layton, I. and J.~Oppenheim (2023).
\newblock The classical-quantum limit.
\newblock {\em PRX Quantum} 5, 020331 
 \newblock \href{https://doi.org/10.1103/PRXQuantum.5.020331}{DOI: 10.1103/PRXQuantum.5.020331}

\bibitem[\protect\citeauthoryear{Leonhardt}{Leonhardt}{2010}]{leonhardt:2010}
Leonhardt, U. (2010).
\newblock {\em Essential quantum optics: from quantum measurements to black
  holes}.
\newblock Cambridge University Press.
 \newblock \href{https://doi.org/10.1017/CBO9780511806117}{DOI: 10.1017/CBO9780511806117}

\bibitem[\protect\citeauthoryear{Lieb and Loss}{Lieb and
  Loss}{2001}]{lieb:2001}
Lieb, E.~H. and M.~Loss (2001).
\newblock {\em Analysis}, Volume~14.
\newblock American Mathematical Soc.
 \newblock \href{https://doi.org/10.1090/gsm/014}{DOI: 10.1090/gsm/014}

\bibitem[\protect\citeauthoryear{March}{March}{2023}]{march:2023}
March, E. (2023).
\newblock Is the Deutsch-Wallace theorem redundant?
\newblock {\em Philosophy of Physics} 7
 \newblock \href{https://doi.org/10.31389/pop.65}{DOI: 10.31389/pop.65}

\bibitem[\protect\citeauthoryear{Mari{\~n}o}{Mari{\~n}o}{2021}]{marino:2021}
Mari{\~n}o, M. (2021).
\newblock {\em Advanced topics in quantum mechanics}.
\newblock Cambridge University Press.
 \newblock \href{https://doi.org/10.1017/9781108863384}{DOI: 10.1017/9781108863384}

\bibitem[\protect\citeauthoryear{Mulder}{Mulder}{2024}]{mulder:2024}
Mulder, R. (2024).
\newblock The classical stance: Dennett's criterion in Wallacian quantum
  mechanics.
\newblock {\em Studies in History and Philosophy of Science\/}~{\em 107},
  11--24.
   \newblock \href{https://doi.org/10.1016/j.shpsa.2024.06.005}{DOI: 10.1016/j.shpsa.2024.06.005}

\bibitem[\protect\citeauthoryear{O'Connell and Wigner}{O'Connell and
  Wigner}{1981}]{OConnell:1981}
O'Connell, R. and E.~Wigner (1981).
\newblock Quantum-mechanical distribution functions: Conditions for uniqueness.
\newblock {\em Physics Letters A\/}~{\em 83\/}(4).
 \newblock \href{https://doi.org/10.1016/0375-9601(81)90870-7}{DOI: 10.1016/0375-9601(81)90870-7}

\bibitem[\protect\citeauthoryear{Palacios}{Palacios}{2018}]{palacios:2018}
Palacios, P. (2018).
\newblock Had we but world enough, and time... but we don't!: Justifying the
  thermodynamic and infinite-time limits in statistical mechanics.
\newblock {\em Foundations of Physics\/}~{\em 48}, 526--541.
 \newblock \href{https://doi.org/10.1007/s10701-018-0165-0}{DOI: 10.1007/s10701-018-0165-0}

\bibitem[\protect\citeauthoryear{Palacios}{Palacios}{2019}]{palacios:2019}
Palacios, P. (2019).
\newblock Phase transitions: A challenge for intertheoretic reduction?
\newblock {\em Philosophy of Science\/}~{\em 86\/}(4), 612--640.
 \newblock \href{https://doi.org/10.1086/704974 }{DOI: 10.1086/704974 } 

\bibitem[\protect\citeauthoryear{Palacios}{Palacios}{2022}]{palacios:2022}
Palacios, P. (2022).
\newblock {\em Emergence and reduction in physics}.
\newblock Cambridge University Press.
 \newblock \href{https://doi.org/10.1017/9781108901017 }{DOI: 10.1017/9781108901017 }

\bibitem[\protect\citeauthoryear{Pathria and Beale}{Pathria and
  Beale}{2011}]{pathria:2011}
Pathria, R. and P.~Beale (2011).
\newblock {\em Statistical Mechanics\/} (Third ed.).
\newblock Elsevier.
 \newblock \href{http://dx.doi.org/10.1016/B978-0-12-382188-1.00015-3}{DOI: 10.1016/B978-0-12-382188-1.00015-3}

\bibitem[\protect\citeauthoryear{Pitowsky}{Pitowsky}{1989}]{pitowsky:1989}
Pitowsky, I. (1989).
\newblock {\em Quantum probability-quantum logic}, Volume 321.
\newblock Springer.
 \newblock \href{http://dx.doi.org/10.1007/BFb0021186}{DOI: 10.1007/BFb0021186}

\bibitem[\protect\citeauthoryear{{Price}}{{Price}}{2010}]{Price:2008}
{Price}, H. (2010).
\newblock {Decisions, Decisions, Decisions: Can Savage Salvage Everettian
  Probability?}
\newblock In S.~Saunders, J.~Barrett, A.~Kent, and D.~Wallace (Eds.), {\em Many
  Worlds? Everett, Quantum Theory, and Reality}, pp.\  369--391. Oxford
  University Press.
 \newblock \href{https://doi.org/10.1093/acprof:oso/9780199560561.001.0001}{DOI: 10.1093/acprof:oso/9780199560561.001.0001}

\bibitem[\protect\citeauthoryear{Rae}{Rae}{2009}]{Rae:2009}
Rae, A.~I. (2009).
\newblock {Everett and the Born rule}.
\newblock {\em Studies In History and Philosophy of Science Part B: Studies In
  History and Philosophy of Modern Physics\/}~{\em 40\/}(3), 243 -- 250.
   \newblock \href{https://doi.org/10.1016/j.shpsb.2009.06.001}{DOI: 10.1016/j.shpsb.2009.06.001}

\bibitem[\protect\citeauthoryear{Read}{Read}{2018}]{read:2018}
Read, J. (2018).
\newblock In defence of Everettian decision theory.
\newblock {\em Studies in History and Philosophy of Science Part B: Studies in
  History and Philosophy of Modern Physics\/}~{\em 63}, 136--140.
   \newblock \href{https://doi.org/10.1016/j.shpsb.2018.01.005}{DOI: 10.1016/j.shpsb.2018.01.005}

\bibitem[\protect\citeauthoryear{Rohrlich}{Rohrlich}{1989}]{rohrlich:1989}
Rohrlich, F. (1989).
\newblock The logic of reduction: The case of gravitation.
\newblock {\em Foundations of Physics\/}~{\em 19\/}(10), 1151--1170.
 \newblock \href{https://doi.org/10.1007/BF00731877}{DOI: 10.1007/BF00731877}

\bibitem[\protect\citeauthoryear{Ross}{Ross}{2000}]{ross:2000}
Ross, D. (2000).
\newblock Rainforest realism: A dennettian theory of existence.
\newblock In D. Ross, D. Thompson \& A. Brook  (Eds.), {\em Dennett's Philosophy: A Comprehensive Assessment }, pp.\  147--168. MIT Press.
 \newblock \href{https://doi.org/10.7551/mitpress/2335.003.0010 }{DOI: 10.7551/mitpress/2335.003.0010 }

\bibitem[\protect\citeauthoryear{Saunders}{Saunders}{2004}]{Saunders:2004}
Saunders, S. (2004).
\newblock Derivation of the Born rule from operational assumptions.
\newblock {\em Proceedings of the Royal Society of London. Series A:
  Mathematical, Physical and Engineering Sciences\/}~{\em 460\/}(2046),
  1771--1788.
   \newblock \href{https://doi.org/10.1098/rspa.2003.1230}{DOI: 10.1098/rspa.2003.1230}

\bibitem[\protect\citeauthoryear{Saunders}{Saunders}{2005}]{Saunders:2005}
Saunders, S. (2005).
\newblock What is probability?
\newblock In E.~Avshalom, S.~Dolev, and N.~Kolenda (Eds.), {\em Quo Vadis
  Quantum Mechanics?}, The Frontiers Collection, pp.\  209--238. Springer.
   \newblock \href{https://doi.org/10.1007/b137897}{DOI: 10.1007/b137897}

\bibitem[\protect\citeauthoryear{Saunders}{Saunders}{2021a}]{saunders:2021b}
Saunders, S. (2021a).
\newblock Branch-counting in the Everett interpretation of quantum mechanics.
\newblock {\em Proceedings of the Royal Society A\/}~{\em 477\/}(2255),
  20210600.
   \newblock \href{https://doi.org/10.1098/rspa.2021.0600}{DOI: 10.1098/rspa.2021.0600}

\bibitem[\protect\citeauthoryear{Saunders}{Saunders}{2024}]{saunders:2024}
Saunders, S. (2024).
\newblock Finite frequentism explains quantum probability.
\newblock {\em British Journal for the Philosophy of Science\/}.
   \newblock \href{https://doi.org/10.1086/731544}{DOI: 10.1086/731544}

\bibitem[\protect\citeauthoryear{Saunders}{Saunders}{2021b}]{Saunders:2021}
Saunders, S.~W. (2021b).
\newblock The Everett interpretation: Probability 1.
\newblock In {\em The Routledge companion to philosophy of physics}, pp.\
  230--246. Routledge.
   \newblock \href{https://doi.org/10.4324/9781315623818 }{DOI: 10.4324/9781315623818 }

\bibitem[\protect\citeauthoryear{Schroeck}{Schroeck}{2013}]{schroeck:2013}
Schroeck, F.~E. (2013).
\newblock {\em Quantum mechanics on phase space}, Volume~74.
\newblock Springer Science \& Business Media.
 \newblock \href{https://doi.org/10.1007/978-94-017-2830-0}{DOI: 10.1007/978-94-017-2830-0}

\bibitem[\protect\citeauthoryear{Short}{Short}{2023}]{Short:2023}
Short, A.~J. (2023, April).
\newblock Probability in many-worlds theories.
\newblock {\em {Quantum}\/}~{\em 7}, 971.
 \newblock \href{https://doi.org/10.22331/q-2023-04-06-971}{DOI: 10.22331/q-2023-04-06-971}

\bibitem[\protect\citeauthoryear{Sneed}{Sneed}{1970}]{sneed:1970}
Sneed, J.~D. (1970).
\newblock Quantum mechanics and classical probability theory.
\newblock {\em Synthese\/}, 34--64.
 \newblock \href{https://doi.org/10.1007/BF00414187}{DOI: 10.1007/BF00414187}

\bibitem[\protect\citeauthoryear{Sorkin}{Sorkin}{1994}]{sorkin:1994}
Sorkin, R.~D. (1994).
\newblock Quantum mechanics as quantum measure theory.
\newblock {\em Modern Physics Letters A\/}~{\em 9\/}(33), 3119--3127.
 \newblock \href{https://doi.org/10.1142/S021773239400294X}{DOI: 10.1142/S021773239400294X}

\bibitem[\protect\citeauthoryear{Sorkin}{Sorkin}{2010}]{sorkin:2010}
Sorkin, R.~D. (2010).
\newblock Logic is to the quantum as geometry is to gravity.
\newblock {\em arXiv preprint arXiv:1004.1226\/}.
 \newblock \href{ 	
https://doi.org/10.48550/arXiv.1004.1226}{ 	
DOI: 10.48550/arXiv.1004.1226}

\bibitem[\protect\citeauthoryear{Steeger}{Steeger}{2022}]{steeger:2022}
Steeger, J. (2022).
\newblock One world is (probably) just as good as many.
\newblock {\em Synthese\/}~{\em 200\/}(2), 97.
 \newblock \href{https://doi.org/10.1007/s11229-022-03499-z}{DOI: 10.1007/s11229-022-03499-z}

\bibitem[\protect\citeauthoryear{Stoica}{Stoica}{2021}]{stoica:2021}
Stoica, O.~C. (2021).
\newblock Standard quantum mechanics without observers.
\newblock {\em Physical Review A\/}~{\em 103\/}(3), 032219.
 \newblock \href{https://doi.org/10.1103/PhysRevA.103.032219}{DOI: 10.1103/PhysRevA.103.032219}

\bibitem[\protect\citeauthoryear{Suppe}{Suppe}{1971}]{suppe:1971}
Suppe, F. (1971).
\newblock On partial interpretation.
\newblock {\em The Journal of Philosophy\/}~{\em 68\/}(3), 57--76.
 \newblock \href{https://doi.org/10.2307/2025168
}{DOI: 10.2307/2025168
}

\bibitem[\protect\citeauthoryear{Suppes}{Suppes}{1961}]{suppes:1961}
Suppes, P. (1961).
\newblock Probability concepts in quantum mechanics.
\newblock {\em Philosophy of Science\/}~{\em 28\/}(4), 378--389.
 \newblock \href{https://doi.org/10.1086/287824 }{DOI: 10.1086/287824 }

\bibitem[\protect\citeauthoryear{Suppes and Zanotti}{Suppes and
  Zanotti}{1993}]{suppes:1993}
Suppes, P. and M.~Zanotti (1993).
\newblock When are probabilistic explanations possible?
\newblock {\em Models and methods in the philosophy of science: selected
  essays\/}, 141--148.
   \newblock \href{https://doi.org/10.1007/978-94-017-2300-8}{DOI: 10.1007/978-94-017-2300-8}

\bibitem[\protect\citeauthoryear{Umekawa, Lee, and Hatano}{Umekawa
  et~al.}{2024}]{umekawa:2024}
Umekawa, S., J.~Lee, and N.~Hatano (2024).
\newblock Advantages of the Kirkwood--Dirac distribution among general
  quasi-probabilities on finite-state quantum systems.
\newblock {\em Progress of Theoretical and Experimental Physics\/}~{\em
  2024\/}(2), 023A02.
   \newblock \href{https://doi.org/10.1093/ptep/ptae005}{DOI: 10.1093/ptep/ptae005}

\bibitem[\protect\citeauthoryear{Wallace}{Wallace}{2002}]{Wallace:2002}
Wallace, D. (2002).
\newblock Quantum probability and decision theory, revisited.
\newblock {\em ArXiv e-prints arXiv:quant-ph/0211104\/}.
 \newblock \href{ 	
https://doi.org/10.48550/arXiv.quant-ph/0211104}{ 	
DOI: 10.48550/arXiv.quant-ph/0211104}

\bibitem[\protect\citeauthoryear{Wallace}{Wallace}{2009}]{Wallace:2009}
Wallace, D. (2009).
\newblock A formal proof of the Born rule from decision-theoretic assumptions.
\newblock {\em ArXiv e-prints http://arxiv.org/abs/0906.2718v1\/}.
 \newblock \href{ 	
https://doi.org/10.48550/arXiv.0906.2718
}{ 	
DOI: 10.48550/arXiv.0906.2718
}

\bibitem[\protect\citeauthoryear{Wallace}{Wallace}{2010}]{Wallace:2010}
Wallace, D. (2010).
\newblock Decoherence and ontology: Or: How i learned to stop worrying and love
  FAPP.
\newblock In S.~Saunders, J.~Barrett, A.~Kent, and D.~Wallace (Eds.), {\em Many
  Worlds? Everett, Quantum Theory, and Reality}, pp.\  53--72. Oxford
  University Press.
 \newblock \href{https://doi.org/10.1093/acprof:oso/9780199560561.001.0001}{DOI: 10.1093/acprof:oso/9780199560561.001.0001}

\bibitem[\protect\citeauthoryear{Wallace}{Wallace}{2012}]{Wallace:2012}
Wallace, D. (2012).
\newblock {\em The Emergent Multiverse}.
\newblock Oxford Univeristy Press.
 \newblock \href{https://doi.org/10.1093/acprof:oso/9780199546961.001.0001}{DOI: 10.1093/acprof:oso/9780199546961.001.0001}

\bibitem[\protect\citeauthoryear{Wallace}{Wallace}{2014}]{Wallace:2014}
Wallace, D. (2014).
\newblock Probability in physics: Statistical, stochastic, quantum.
\newblock {\em Chance and Temporal Asymmetry, edited by Alastair Wilson. Oxford
  University Press\/}.
   \newblock \href{https://doi.org/10.1093/acprof:oso/9780199673421.001.0001}{DOI: 10.1093/acprof:oso/9780199673421.001.0001}

\bibitem[\protect\citeauthoryear{Wallace}{Wallace}{2021}]{Wallace:2021}
Wallace, D. (2021).
\newblock Probability and irreversibility in modern statistical mechanics:
  Classical and quantum.
\newblock {\em arXiv preprint arXiv:2104.11223\/}.
 \newblock \href{ 	
https://doi.org/10.48550/arXiv.2104.11223}{ 	
DOI: 10.48550/arXiv.2104.11223}

\bibitem[\protect\citeauthoryear{Wigner}{Wigner}{1932}]{Wigner:1932}
Wigner, E. (1932).
\newblock On the quantum correction for thermodynamic equilibrium.
\newblock {\em Physical review\/}~{\em 40\/}(5), 749.
 \newblock \href{https://doi.org/10.1103/PhysRev.40.749}{DOI:10.1103/PhysRev.40.749}

\bibitem[\protect\citeauthoryear{Wigner}{Wigner}{1971}]{Wigner:1971}
Wigner, E.~P. (1971).
\newblock Quantum-mechanical distribution functions revisited.
\newblock In {\em Part I: Physical Chemistry. Part II: Solid State Physics},
  pp.\  251--262. Springer.
   \newblock \href{https://doi.org/10.1007/978-3-642-59033-7}{DOI: 10.1007/978-3-642-59033-7}

\bibitem[\protect\citeauthoryear{Zurek}{Zurek}{2005}]{Zurek:2005}
Zurek, W.~H. (2005, May).
\newblock Probabilities from entanglement, {Born}'s rule
  ${p}_{k}=\mid\psi_{k}\mid^{2}$ from envariance.
\newblock {\em Phys. Rev. A\/}~{\em 71}, 052105.
 \newblock \href{10.1103/PhysRevA.71.052105}{DOI: 10.1103/PhysRevA.71.052105}

\end{thebibliography}
\end{document}